\documentclass[twocolumn]{aastex701}

\usepackage[T1]{fontenc} 
\usepackage[utf8]{inputenc} 
\usepackage{amssymb}
\usepackage{float}
\usepackage{placeins}
\usepackage{hyperref}
\usepackage{natbib}
\usepackage{lipsum}

\usepackage{subcaption}

\setcitestyle{notesep={ }}

\begin{document}

\title{POSEIDON III: The Aligned Orbit of the Hot Neptune Around the Hot Star WASP-195}

\correspondingauthor{Juan I.\ Espinoza-Retamal}
\email{jiespinozar@princeton.edu}

\author[0000-0001-9480-8526]{Juan I.\ Espinoza-Retamal}
\altaffiliation{51 Pegasi b Fellow}
\affiliation{Department of Astrophysical Sciences, Princeton University, 4 Ivy Lane, Princeton, NJ 08540, USA}
\email{jiespinozar@princeton.edu}

\author[0000-0002-4265-047X]{Joshua N.\ Winn}
\affiliation{Department of Astrophysical Sciences, Princeton University, 4 Ivy Lane, Princeton, NJ 08540, USA}
\email{jnwinn@princeton.edu}

\author[0000-0002-9158-7315]{Rafael Brahm}
\affiliation{Facultad de Ingenier\'ia y Ciencias, Universidad Adolfo Ib\'{a}\~{n}ez, Av.\ Diagonal Las Torres 2640, 7941169 Pe\~{n}alol\'{e}n, Santiago, Chile}
\email{rafael.brahm@uai.cl}

\author[0000-0002-9305-5101]{Luke B.\ Handley}
\altaffiliation{NSF Graduate Research Fellow}
\affil{Department of Astronomy, California Institute of Technology, Pasadena, CA 91125, USA}
\email{lhandley@caltech.edu}

\author[0009-0007-0740-0954]{Elise Koo}
\affiliation{Anton Pannekoek Institute for Astronomy, University of Amsterdam, Science Park 904, 1098 XH Amsterdam, The Netherlands}
\affiliation{ASTRON, Netherlands Institute for Radio Astronomy, Oude Hoogeveensedijk 4, Dwingeloo 7991 PD, The Netherlands}
\email{e.j.m.koo@uva.nl}

\author[0000-0001-9985-0643]{Caleb Lammers}
\affiliation{Department of Astrophysical Sciences, Princeton University, 4 Ivy Lane, Princeton, NJ 08540, USA}
\email{caleb.lammers@princeton.edu}

\author[0000-0003-0412-9314]{Cristobal Petrovich}
\affiliation{Department of Astronomy, Indiana University, 727 East 3rd Street, Bloomington, IN 47405, USA}
\email{cpetrovi@iu.edu}

\author[0000-0001-7409-5688]{Guðmundur Stefánsson}
\affiliation{Astrophysics \& Space Center, Schmidt Sciences, New York, NY 10011, USA}
\affiliation{Anton Pannekoek Institute for Astronomy, University of Amsterdam, Science Park 904, 1098 XH Amsterdam, The Netherlands}
\email{g.k.stefansson@uva.nl}

\author[0000-0002-5389-3944]{Andr\'es Jord\'an}
\affiliation{Facultad de Ingenier\'ia y Ciencias, Universidad Adolfo Ib\'{a}\~{n}ez, Av.\ Diagonal Las Torres 2640, 7941169 Pe\~{n}alol\'{e}n, Santiago, Chile}
\affiliation{Departamento de Astronomía, Universidad de Chile, Camino El Observatorio 1515, 7591245 Las Condes, Santiago, Chile}
\affiliation{El Sauce Observatory --- Obstech, Coquimbo, Chile}
\email{andres.jordan@uai.cl}

\author[0000-0002-0376-6365]{Xian-Yu Wang}
\altaffiliation{Sullivan Prize Postdoctoral Fellow}
\affiliation{Department of Astronomy, Indiana University, 727 East 3rd Street, Bloomington, IN 47405, USA}
\email{xwa5@iu.edu}

\author[0000-0002-7846-6981]{Songhu Wang}
\affiliation{Department of Astronomy, Indiana University, 727 East 3rd Street, Bloomington, IN 47405, USA}
\email{sw121@iu.edu}

\author[0000-0003-2657-3889]{Nicholas Saunders}
\affiliation{Department of Astronomy, Yale University, 219 Prospect Street, New Haven, CT 06511, USA}
\email{nicholas.saunders@yale.edu}

\author[0000-0003-0967-2893]{Erik A.\ Petigura}
\affiliation{Department of Physics \& Astronomy, University of California Los Angeles, Los Angeles, CA 90095, USA}
\email{petigura@astro.ucla.edu}

\author[0000-0002-3725-3058]{Lauren M.\ Weiss}
\affiliation{Department of Physics and Astronomy, University of Notre Dame, Notre Dame, IN 46556, USA}
\email{lweiss4@nd.edu}

\author[0000-0002-6525-7013]{Ashley D.\ Baker}
\affiliation{Department of Astronomy, California Institute of Technology, Pasadena, CA 91125, USA}
\email{abaker@caltech.edu}

\author[0000-0001-6416-1274]{Theron W.\ Carmichael}
\altaffiliation{NSF Ascend Postdoctoral Fellow}
\affiliation{Institute for Astronomy, University of Hawai‘i, 2680 Woodlawn Drive, Honolulu, HI 96822, USA}
\email{tcarmich@hawaii.edu}

\author[0000-0002-8958-0683]{Fei Dai}
\affiliation{Institute for Astronomy, University of Hawai‘i, 2680 Woodlawn Drive, Honolulu, HI 96822, USA}
\email{fdai@hawaii.edu}

\author[0009-0002-2419-8819]{Jerry Edelstein}
\affiliation{Space Sciences Laboratory, University of California Berkeley, Berkeley, CA 94720, USA}
\email{jerrye@ssl.berkeley.edu}

\author[0009-0005-0100-7612]{Jack Foley}
\affiliation{Department of Physics \& Astronomy, University of California Los Angeles, Los Angeles, CA 90095, USA}
\email{jackfoley1929@g.ucla.edu}

\author[0000-0003-3504-5316]{Benjamin J.\ Fulton}
\affiliation{NASA Exoplanet Science Institute/Caltech-IPAC, California Institute of Technology, Pasadena, CA 91125, USA}
\email{bjfulton@ipac.caltech.edu}

\author[0000-0002-8965-3969]{Steven Giacalone}
\affiliation{Department of Astronomy, California Institute of Technology, Pasadena, CA 91125, USA}
\email{giacalone@astro.caltech.edu}

\author[0009-0004-4454-6053]{Steven R.\ Gibson}
\affiliation{Caltech Optical Observatories, California Institute of Technology, Pasadena, CA 91125, USA}
\email{sgibson@caltech.edu}

\author[0000-0003-1312-9391]{Samuel Halverson}
\affiliation{Jet Propulsion Laboratory, California Institute of Technology, 4800 Oak Grove Drive, Pasadena, CA 91109, USA}
\email{samuel.halverson@jpl.nasa.gov}

\author[0000-0001-8638-0320]{Andrew W.\ Howard}
\affiliation{Department of Astronomy, California Institute of Technology, Pasadena, CA 91125, USA}
\email{ahoward@caltech.edu}

\author[0000-0002-0531-1073]{Howard Isaacson}
\affiliation{Department of Astronomy, University of California Berkeley, Berkeley, CA 94720, USA}
\email{hisaacson@berkeley.edu}

\author[0009-0006-4095-4097]{Emma McIntyre}
\affiliation{Institute for Astronomy, University of Hawai‘i, 2680 Woodlawn Drive, Honolulu, HI 96822, USA}
\email{edm7@hawaii.edu}

\author[0000-0001-5728-4735]{Pranav H.\ Premnath}
\affiliation{Department of Physics \& Astronomy, University of California Irvine, Irvine, CA 92697, USA}
\email{premnatp@uci.edu}

\author[0009-0008-9238-5871]{Maleah Rhem}
\affiliation{Department of Physics and Astronomy, University of Kansas, Lawrence, KS, USA}
\email{mkrhem@ku.edu}

\author{Kodi Rider}
\affiliation{Space Sciences Laboratory, University of California Berkeley, Berkeley, CA 94720, USA}
\email{kodi.rider@ssl.berkeley.edu}

\author[0000-0003-3856-3143]{Ryan A.\ Rubenzahl}
\affiliation{Center for Computational Astrophysics, Flatiron Institute, 162 Fifth Avenue, New York, NY 10010, USA}
\email{rrubenzahl@gmail.com}

\author[0000-0001-8127-5775]{Arpita Roy}
\affiliation{Astrophysics \& Space Center, Schmidt Sciences, New York, NY 10011, USA}
\email{arpita308@gmail.com}

\author[0009-0004-7325-3591]{Christopher Smith}
\affiliation{Space Sciences Laboratory, University of California Berkeley, Berkeley, CA 94720, USA}
\email{christopher.smith@berkeley.edu}

\author[0000-0002-4290-6826]{Judah Van Zandt}
\affiliation{Department of Physics, University of California Santa Barbara, Santa Barbara, CA 93106, USA}
\email{judahvz@ucsb.edu}

\author[0000-0002-6092-8295]{Josh Walawender}
\affiliation{W.\ M.\ Keck Observatory, 65-1120 Mamalahoa Hwy, Kamuela, HI 96743, USA}
\email{jwalawender@keck.hawaii.edu}

\author[0009-0004-0455-2424]{Elina Y.\ Zhang}
\affiliation{Institute for Astronomy, University of Hawai‘i, 2680 Woodlawn Drive, Honolulu, HI 96822, USA}
\email{yuchenzh@hawaii.edu}

\author[0009-0005-7924-3690]{Jingyi Zhang}
\affiliation{Institute for Astronomy, University of Hawai‘i, 2680 Woodlawn Drive, Honolulu, HI 96822, USA}
\email{jzhang97@hawaii.edu}

\begin{abstract}

Stellar obliquities provide important clues as to the formation and migration histories of planetary systems, but measurements remain scarce
for Neptune-mass planets, especially those orbiting hot stars (above the Kraft break). Here we present observations of the Rossiter-McLaughlin effect in the hot-star/hot-Neptune system WASP-195 ($T_{\rm eff}=6470\pm100$~K, $v\sin{i_\star}=10.5\pm1.1$~km~s$^{-1}$) obtained with the Keck Planet Finder and NEID spectrographs. A joint analysis of these observations, archival photometry, and archival radial velocities yields a sky-projected stellar obliquity of $\lambda=-10\pm7^\circ$, consistent with spin-orbit alignment. This makes WASP-195 one of the few hot-star/hot-Neptune systems with a measured obliquity. Archival radial velocities from SOPHIE exclude Jupiter-mass planets within approximately 3~au at $5\sigma$ confidence. The aligned and nearly circular orbit is naturally consistent with a history of disk-driven migration, although coplanar high-eccentricity migration or Roche-lobe overflow cannot be ruled out. We also investigate why so few Neptunes around hot stars have measured obliquities. Their scarcity likely reflects a combination of the lower intrinsic occurrence of short-period Neptunes around hot stars and the difficulty of confirming planet candidates in this regime, where rapid stellar rotation broadens spectral lines and hampers conventional radial-velocity confirmation. Rapid rotation also increases the detectability of the Rossiter-McLaughlin effect, a feature that could help to widen the planet confirmation bottleneck while expanding the obliquity census of small planets around hot stars.

\end{abstract}

\section{Introduction}

The orbital paths of the Solar System planets are nearly coplanar, with a median inclination of $\sim1^{\circ}$ with respect to the invariable plane, which is itself only 6$^\circ$ from
the Sun's equatorial plane \citep[e.g.,][]{Beck2005,Souami2012S}. The high degree of coplanarity motivated the early nebular hypotheses of Kant and Laplace, who in the 18th century proposed that the Solar System formed from a gas cloud that collapsed to become the Sun, surrounded by a disk of material from which planets formed. In this picture, coplanarity is a natural outcome. Today, of course, flat and circular protoplanetary disks are routinely observed around young stars \citep[e.g.,][]{Andrews2020,Kurtovic2026}.

However, the discovery of thousands of exoplanets has revealed that planetary systems do not always have such orderly architectures. For systems with transiting planets, one way to probe their geometries is through the stellar obliquity $\psi$, defined as the angle between the stellar spin axis and the planet's orbital axis. Stellar obliquities can be constrained using observations of the Rossiter--McLaughlin (RM) effect \citep{Rossiter1924,McLaughlin1924}, which measures the sky-projected obliquity $\lambda$ \citep[e.g.,][]{Triaud2018, Albrecht2022}. While most planetary systems are well aligned, a notable population exhibits large spin--orbit misalignments, demonstrating that the coplanarity observed in the Solar System is not universal \citep[e.g.,][]{Hebrard2008,Hjorth2021,Stefansson2022,Weldon2026}.

Most stellar obliquity measurements have been obtained for hot Jupiters. Since these planets are thought to form at much larger orbital separations \citep[e.g.,][]{Dawson2018,He2026}, their obliquities have long been interpreted as tracers of the migration mechanisms that brought them close to their host stars \citep[e.g.,][]{Queloz2000,Winn2005}. However, the interpretation of these obliquities remains uncertain. Tidal dissipation of oscillations induced by hot Jupiters on their host stars may damp stellar obliquities over time, particularly for cool stars \citep[e.g.,][]{Schlaufman2010,Winn2010,Albrecht2012}, complicating the interpretation of their observed architectures.

Neptunes represent a distinct and more common population of short-period planets, \added{occurring around approximately $6.5\%$ of Sun-like stars with orbital periods $P<50$~days} \citep[e.g.,][]{Howard2010science}. Measuring their stellar obliquities provides an opportunity to probe the formation and dynamical evolution of Neptunes and to determine whether their architectures resemble those observed among hot Jupiters. Although recent observations have substantially expanded the number of Neptunes with measured stellar obliquities \citep[e.g.,][]{Espinoza-Retamal2024,Bourrier2025,Handley2026}, the sample remains small, particularly around hot stars \citep[e.g.,][]{Dugan2025,Espinoza-Retamal2026}. A larger and more diverse sample is therefore needed to characterize the obliquity distribution of Neptune-mass planets and identify possible dependencies on planetary and stellar properties.

In this paper, the third in the POSEIDON series \citep{Espinoza-Retamal2026,Espinoza-Retamal2026b}, we investigate the architecture of WASP-195 \citep{Schanche2025}, a hot star with $T_{\rm eff}=6470\pm100$~K that lies above the
\cite{Kraft1967} break at $\approx6250$~K and hosts a hot Neptune with a 5-day orbit and a low bulk density ($\rho_p \sim 0.2$~g~cm$^{-3}$, $M_p\sim36\,M_\oplus$, $R_p\sim9.5\,R_\oplus$). We present observations of the RM effect to measure the sky-projected stellar obliquity of the system, together with photometric and spectroscopic observations to refine its planetary and orbital properties and search for additional companions. 

We have structured this paper as follows. Section~\ref{sec:observations} describes the new and archival observations of WASP-195. Section~\ref{sec:stellar} presents the stellar characterization, and Sections~\ref{sec:phot} and \ref{sec:fit} describe the photometric analysis and the joint analysis for the obliquity measurement, respectively. In Section~\ref{sec:discussion}, we discuss our results in the context of the broader population of transiting Neptunes, and the architecture and possible dynamical history of the WASP-195 system. Finally, Section~\ref{sec:conclusion} summarizes our findings and conclusions.

\section{Observations}\label{sec:observations}

\subsection{NEID Transit Spectroscopy}

We observed a transit of WASP-195~b using the NEID spectrograph \citep{Schwab2016}, installed on the WIYN 3.5\,m telescope at Kitt Peak Observatory in Arizona (Program ID 2025A-944383, PI Songhu Wang). NEID is an environmentally stabilized, fiber-fed echelle spectrograph covering the wavelength range 380--930~nm with a resolving power of $R\approx110,\!000$ \citep{Stefansson2016,Robertson2019,Kanodia2018,Halverson2016}. The transit was observed on UTC 2025 May 17, between 03:31 and 08:42. We obtained 17 spectra with an exposure time of 1100~s. The data were processed with the dedicated NEID data reduction pipeline \citep{Bender2022}, which resulted in final spectra with a median signal-to-noise ratio (S/N) of 17 per extracted pixel at 550~nm. We extracted precise radial velocities (RVs) using the \texttt{serval} template-matching code \citep{Zechmeister2018} adapted for NEID by \citet{Stefansson2022}. The resulting RVs have a median uncertainty of 17~m~s$^{-1}$, derived from the order-by-order uncertainties. The NEID RVs, along with the best model, are shown in Figure~\ref{fig:fit} and are available as the data behind the figure.

\subsection{KPF Transit Spectroscopy}

\begin{figure*}[t!]
    \centering
    \includegraphics[width=\linewidth]{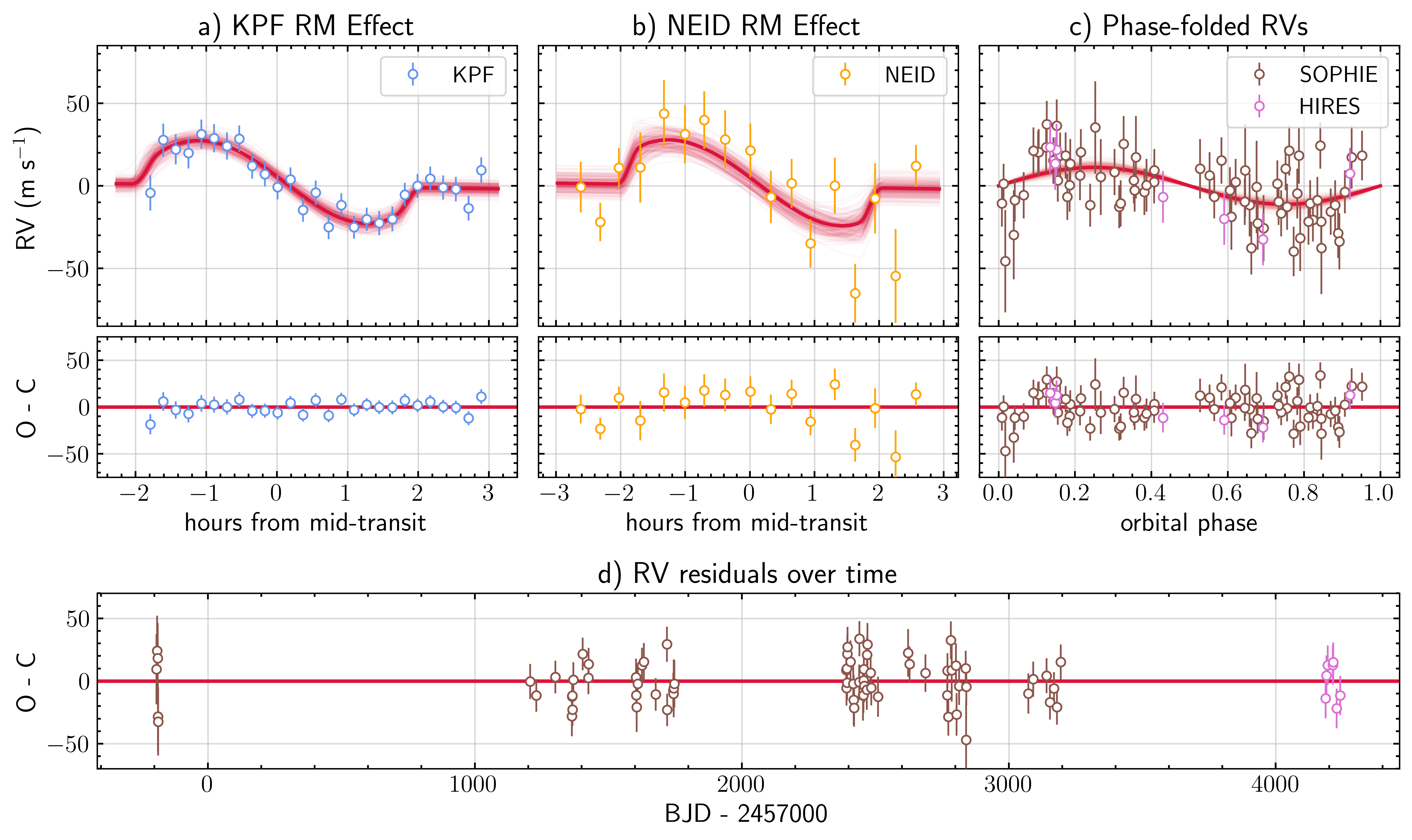}
    \includegraphics[width=0.7\linewidth]{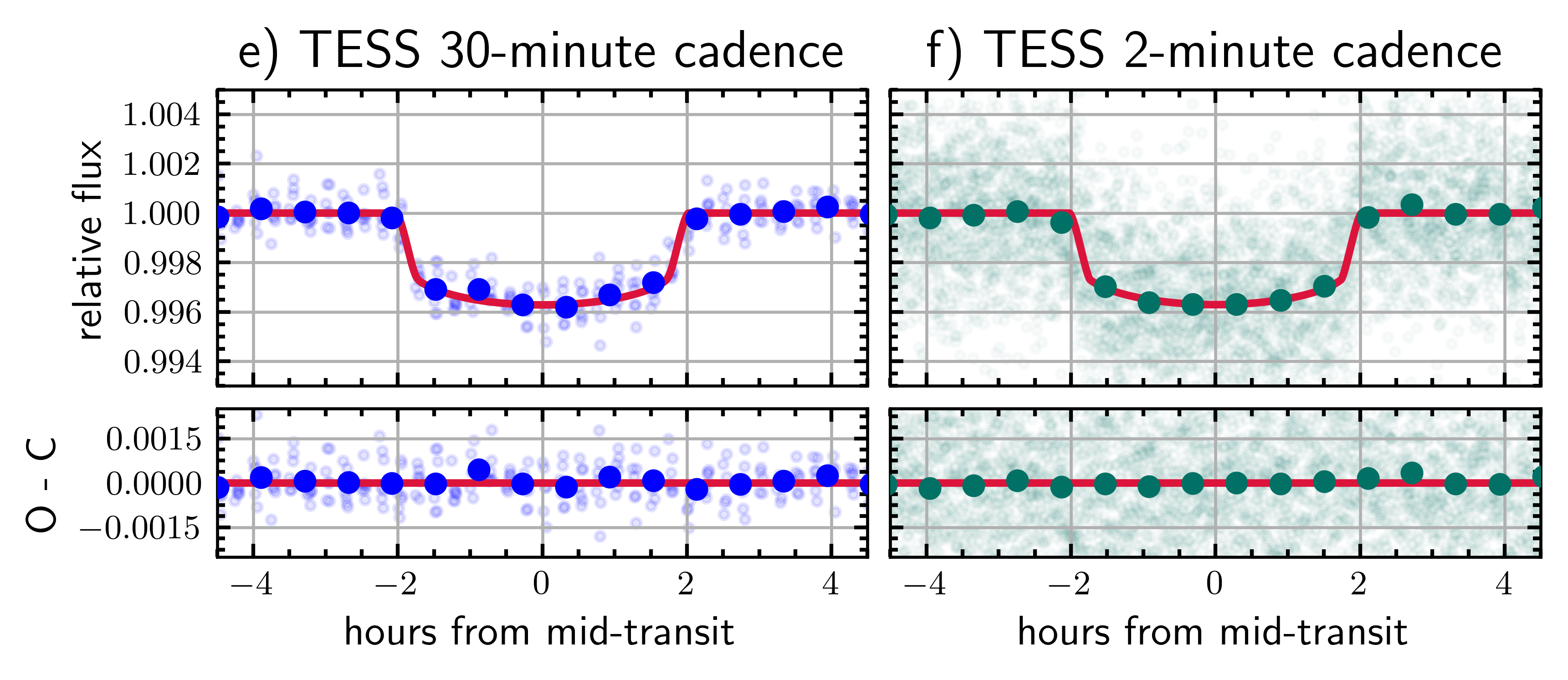}
    \caption{Radial-velocity and photometric observations of WASP-195. In all cases, the red curves are best-fit models, and the lighter red curves are models drawn randomly from the posteriors. The residuals are plotted beneath the data. The RV error bars include a white noise jitter term added in quadrature. a) KPF velocities spanning a transit. b) NEID velocities spanning a transit. c) SOPHIE and HIRES velocities across all orbital phases. d) SOPHIE and HIRES RV residuals as a function of time. e) Phase-folded photometry based on TESS observations with 30-minute cadence (blue). The larger and darker points are time-averaged data. f) Phase-folded photometry based on TESS observations with 2-minute cadence (green). The larger and darker points are time-averaged data. KPF, NEID, and HIRES RVs are available electronically, as the data behind the figure.}
    \label{fig:fit}
\end{figure*}

We observed a second transit of WASP-195~b using the Keck Planet Finder (KPF) spectrograph \citep{Gibson2018,Gibson2020,Gibson2024}, mounted on the Keck~I 10~m telescope at Mauna Kea Observatory in Hawaii (Program ID 2026A\_N095, PI Juan I.\ Espinoza-Retamal). KPF is a stabilized, fiber-fed echelle spectrograph covering the wavelength range 445--870~nm at a resolving power of $R\approx100,\!000$ \citep{Lilley2022,Sirk2018}. The observations were carried out on UTC 2026 March 31, between 10:51 and 15:32. In total, we obtained 27 spectra of the host star with an exposure time of 600~s each. Conditions were clear throughout the sequence, with a median seeing of $0.43^{\prime\prime}$, estimated from Moffat-profile fits to the guider images. The airmass decreased from 1.59 to a minimum of 1.15 and ended at 1.21. The spectra were processed with the standard KPF data reduction pipeline \citep{Gibson2020}, which also delivered RVs using the cross-correlation function method. The final spectra have a median S/N of 90 per extracted pixel at 550~nm, resulting in a median RV uncertainty of 7.7~m~s$^{-1}$. KPF RVs of WASP-195, along with the best model, are shown in Figure~\ref{fig:fit} and are available as the data behind the figure.

\subsection{HIRES Spectroscopy}

In addition to our in-transit observations, we also observed WASP-195 outside of transits using the High Resolution Echelle Spectrometer \citep[HIRES;][]{Vogt1994} installed on the Keck~I 10~m telescope at Mauna Kea Observatory in Hawaii. Observations were executed through the California Planet Search \citep{Howard2010} queue system at the observatory, which employs the same optimization algorithms as the KPF-Community Cadence scheduler \citep{Lubin2026,Handley2024b,Handley2024a}. HIRES is an echelle spectrograph that covers the wavelength range 374--970~nm at a resolving power of $\sim$ 25,000--85,000 depending on the slit used. We obtained nine spectra of the star between May and July 2026 with exposure times of $\sim680$~s. In order to derive precise RVs, we also obtained an iodine-free template spectrum on the night of July 21, 2026, with an exposure time of $\sim1300$~s and a narrower slit. The spectra were processed and RVs were extracted using the standard HIRES procedure described by \citet{Howard2010}, which relies on the iodine-cell method of \citet{Butler1996}. The HIRES RVs have a median formal uncertainty of 4.3~m~s$^{-1}$ as estimated from the scatter among the fits to individual spectral segments spanning a few Angstroms. The HIRES RVs, along with the best model, are shown in Figure~\ref{fig:fit} and are available as the data behind the figure.

\subsection{TESS Photometry}

Since the discovery of the planet by \citet{Schanche2025}, additional photometric observations of WASP-195 have been collected by the Transiting Exoplanet Survey Satellite \citep[TESS;][]{Ricker2015}. We used the TESS data to refine the system ephemeris and to jointly model the light curves with the spectroscopic observations. The TESS light curves were retrieved from the Mikulski Archive for Space Telescopes using the \texttt{lightkurve} package \citep{lightkurve}. In particular, we used the 30-minute cadence light curves \citep{TESS_30min} from Sectors 23 and 25, and the 2-minute cadence light curves \citep{TESS_2min} from Sectors 50, 52, 78, and 79. All light curves were processed with the TESS Science Processing Operations Center pipeline \citep{spoc}, which corrects for instrumental systematics including pointing and focus variations, discontinuities caused by radiation events in the CCD detectors, outliers, and flux contamination from nearby stars. The TESS transit observations of WASP-195, along with the best-fitting models, are shown in Figure~\ref{fig:fit}.

\subsection{Archival Spectroscopy}

In addition to our new RV observations, in the analysis we included 73 out-of-transit RV measurements of WASP-195 obtained by \citet{Schanche2025} between May 2014 and September 2023. These observations were taken with the SOPHIE spectrograph mounted on the 1.93~m telescope at the Observatoire de Haute-Provence in France
\citep{Perruchot2008,Bouchy2009}, and were originally used to confirm the planetary nature of the transiting object and measure its mass. In this work, we jointly model these archival RVs together with our new observations in order to refine the orbital and planetary parameters. The archival SOPHIE observations of WASP-195, along with the best-fit model, are shown in Figure~\ref{fig:fit}.

\section{Stellar Parameters}\label{sec:stellar}

As part of the POSEIDON survey, we are performing a homogeneous analysis of the survey targets, including uniformly derived stellar parameters. We have adopted the method described by \citet{Brahm2019} and applied it here to derive stellar parameters of WASP-195 based on the HIRES iodine-free template spectrum. Briefly, the method consists of two steps that are repeated iteratively. In the first step, we compute the stellar atmospheric parameters using the \texttt{zaspe} package \citep{zaspe}, which compares the high-resolution spectrum with a grid of synthetic spectra to identify the closest match. The search is performed in the spectral regions most sensitive to changes in the stellar parameters, and error bars are computed through Monte Carlo simulations. In the second step, we compute the stellar physical parameters by fitting stellar evolutionary models to the observed spectral energy distribution. We fit the available broadband apparent magnitudes and the parallax from Gaia Data Release 3 \citep[DR3;][]{GaiaDR3} to synthetic magnitudes generated from the \texttt{PARSEC} isochrones \citep{parsec} and the \citet{Cardelli1989} model for interstellar extinction. In this step, the stellar temperature derived with \texttt{zaspe} is used as a prior and the metallicity is held fixed. From the stellar mass and radius obtained with the second step, we calculate a more precise $\log{g}$, which is held fixed in a new iteration of the first step. We iterated the two procedures until reaching convergence, which happens when two consecutive \texttt{zaspe} runs agree on the best-fitting grid values of $T_{\rm eff}$ and [Fe/H]. 
The derived stellar parameters of WASP-195 are consistent with those reported in the literature and are presented in Table~\ref{tab:stellar}.

\begin{deluxetable*}{llcr}
\tablecaption{Stellar properties of WASP-195.\label{tab:stellar}}
\tablecolumns{4}
\tablewidth{0pt}
\tablehead{Parameter & Description & Value & Reference}
\startdata
RA & Right Ascension (J2015.5) & 16h30m11.91s & \citet{GaiaDR3}\\
Dec & Declination (J2015.5) & +49d53m44.85s & \citet{GaiaDR3}\\
pm$^{\rm RA}$ & Proper motion in RA (mas yr$^{-1}$) & $-3.892\pm0.012$ & \citet{GaiaDR3}\\
pm$^{\rm Dec}$ & Proper motion in DEC (mas yr$^{-1}$) & $14.546\pm0.014$ & \citet{GaiaDR3}\\
$\pi$ & Parallax (mas) & $2.03\pm0.01$ & \citet{GaiaDR3} \\
$d$ & Distance (pc) & $493.4\pm2.5$ & \citet{GaiaDR3} \\
\hline
T & TESS magnitude (mag) & $11.435\pm0.007$ & \citet{Stassun2018,Stassun2019}\\
B & B-band magnitude (mag) & $12.3\pm0.3$ & \citet{apass}\\
V & V-band magnitude (mag) & $11.95\pm0.02$ & \citet{apass}\\
G & Gaia G-band magnitude (mag) & $11.7822\pm0.0002$ & \citet{GaiaDR3}\\
G$_{\rm BP}$ & Gaia BP-band magnitude (mag) & $12.0350\pm0.0006$ & \citet{GaiaDR3}\\
G$_{\rm RP}$ & Gaia RP-band magnitude (mag) & $11.3742\pm0.0003$ &  \citet{GaiaDR3}\\
J & 2MASS J-band magnitude (mag) & $10.93\pm0.02$ & \citet{2mass}\\
H & 2MASS H-band magnitude (mag) & $10.70\pm0.02$ & \citet{2mass}\\
K$_s$ & 2MASS K$_s$-band magnitude (mag) & $10.68\pm0.02$ & \citet{2mass}\\
\hline
$T_{\rm eff}$ & Effective temperature (K) & $6470\pm100$ & This work\\
$\log{g}$ & Surface gravity (with $g$ in cm\,s$^{-2}$) & $4.20\pm0.01$ & This work\\
$[$Fe/H$]$ & Metallicity (dex) & $+0.12\pm0.05$ & This work\\
$v\sin{i_\star}$ & Projected rotational velocity (km s$^{-1}$) & $8.9\pm0.5$ & This work\\
$M_{\star}$ & Mass ($M_\odot$) & $1.32\pm0.02$ & This work\\
$R_{\star}$ & Radius ($R_\odot$) & $1.51\pm0.01$ & This work\\
$L_{\star}$ & Luminosity ($L_\odot$) & $3.6\pm0.2$ & This work\\
$A_{V}$ & Visual extinction (mag) & $0.16\pm0.06$ & This work\\
Age & Age (Gyr) & $2.1\pm0.4$ & This work\\
$\rho_\star$ & Mean density (g cm$^{-3}$) & $0.54\pm0.02$ & This work\\
\enddata
\tablecomments{The uncertainties do not take into account possible systematic differences among different stellar evolutionary models \citep{Tayar2022}. The TESS magnitude is shown only for reference and was not included in our stellar analysis.}
\end{deluxetable*}

\section{Photometric Analysis}\label{sec:phot}

In order to refine the orbital ephemeris and to look for possible transit timing variations (TTVs) that might reveal the presence of additional planets in the system, we analyzed the TESS data with the \texttt{juliet} code \citep{juliet}. We modeled the transits with \texttt{batman} \citep{batman} and included a Matern-3/2 Gaussian Process (GP) with \texttt{celerite} \citep{celerite} to remove long-term variability and systematic noise. We used \texttt{dynesty}'s dynamic nested sampler \citep{dynesty2} for sampling the posterior probability distributions.

We placed uniform priors on the impact parameter $b$ and radius ratio $R_p/R_{\star}$, with an informative Gaussian prior on the stellar density $\rho_\star$ based on estimates of the stellar mass and radius described in Section \ref{sec:stellar}. We adopted a quadratic limb darkening law and sampled the coefficients $q_1$ and $q_2$ defined by \citet{Kipping13} with uniform priors. We placed broad uniform priors on the time of each transit midpoint, within 1~day of the expected values calculated from the orbital period and time of mid-transit from \citet{Schanche2025}. We considered the 30-minute and 2-minute TESS light curves as coming from two different instruments, each with its own GP kernel to account for differences in variability captured in different epochs and cadences, while sharing the limb darkening coefficients.

From this analysis, we ruled out TTVs larger than 30 minutes over the observational baseline of 4 years and obtained updated transit ephemerides consistent with the values reported by \citet{Schanche2025}. The uncertainty in the transit midpoint is reduced by roughly a factor of two, while the orbital period is constrained to a similar precision to that of \citet{Schanche2025}. Additionally, we constructed detrended TESS light curves that were used in the global modeling described below.

\section{Obliquity Determination}\label{sec:fit}

In order to derive the stellar obliquity, we followed the same procedure as in previous POSEIDON papers \citep{Espinoza-Retamal2026,Espinoza-Retamal2026b}. In brief, we used the \texttt{ironman} code \citep{Espinoza-Retamal2023b,Espinoza-Retamal2024} to jointly model all the observations described in Section~\ref{sec:observations}. This code uses \texttt{batman} \citep{batman} to model the transit light curves, \texttt{radvel} \citep{Fulton2018} to model the Keplerian RVs, \texttt{rmfit} \citep{Stefansson2022} to model the RM effect, and \texttt{dynesty} \citep{dynesty2} to sample the posteriors. We adopted 4500 live points and the sampler's default configurations for the bounding and sampling methods (i.e., multi-ellipsoidal bounding and random-walk sampling). The runs were stopped when the change in the logarithm of the Bayesian evidence ($\log{Z}$) was less than 0.01, which is the default convergence criterion. Since the stellar rotation period of WASP-195 is not known, the stellar inclination angle cannot be constrained. As a result, the sky-projected stellar obliquity can be determined from the data, but the true obliquity is undetermined.

In this analysis, we only considered the TESS data within 10 hours of the transit midpoint to reduce the computational cost. We excluded two HIRES RV measurements that were taken during the transits of the planet. We included independent jitter terms for each instrument to account for possible stellar activity or instrumental systematics. We placed uniform or log-uniform priors on almost all the parameters (see Table~\ref{tab:fit}). For the orbital period and time of midtransit, we placed uniform priors spanning $\pm10\sigma$ around the values derived in Section~\ref{sec:phot}. More informative priors were used for the stellar mean density $\rho_\star$ and the non-rotational linewidth $\beta$, which accounts for instrumental and macroturbulence broadening \citep[see][]{Hirano2010}. We placed an informative Gaussian prior for $\rho_\star$ based on the value derived in Section \ref{sec:stellar}, while for $\beta$ we considered an instrumental broadening of 3.0~km~s$^{-1}$ and a macroturbulence broadening of 5.6~km~s$^{-1}$ derived from the macroturbulence law for hot stars from \citet{Gray1984}. We added the instrumental and macroturbulence broadening in quadrature to set our prior, with an uncertainty of 2 km s$^{-1}$. All priors and results obtained from the posterior distributions are shown in Table~\ref{tab:fit}.

We found that WASP-195~b has an aligned orbit with a sky-projected spin-orbit angle of $\lambda=-10\pm7^{\circ}$. The data and best-fitting model are shown in Figure~\ref{fig:fit}. We also explored models in which the orbital eccentricity was allowed to vary. However, the difference in $\log{Z}$ between the circular and eccentric models is $\approx2$, indicating that the current data do not provide a statistically significant preference for either solution. As discussed below, tidal dissipation is expected to circularize the orbit on a timescale much shorter than the age of the system. We therefore adopt the circular solution throughout the remainder of this work, while also reporting a $2\sigma$ upper limit of $e<0.2$ obtained from the eccentric fit. The reported parameters are in good agreement with those reported by \citet{Schanche2025}.

\begin{deluxetable*}{llcr}[h!]
\tablecaption{Summary of priors and posteriors of the \texttt{ironman} fit for WASP-195. \label{tab:fit}}
\tablewidth{70pt}
\tablehead{Parameter & Description & Prior & Posterior}
\startdata
$\lambda$ & Sky-projected stellar obliquity (deg) & $\mathcal{U}(-180,180)$ & $-10\pm7$ \\
$v\sin{i_\star}$ & Projected rotational velocity (km s$^{-1}$) & $\mathcal{U}(0,20)$ & $10.5\pm1.1$ \\
$\rho_\star$ & Stellar density (g cm$^{-3}$) & $\mathcal{G}(0.54,0.02)$ & $0.54\pm0.02$ \\
\hline
$P$ & Orbital period (days) & $\mathcal{U}(5.051867,5.051976)$ & $5.051925\pm0.000004$ \\ 
$t_0$ & Transit midpoint (BJD) & $\mathcal{U}(2458928.3776,2458928.3985)$ & $2458928.3876\pm0.0009$ \\
$b$ & Impact parameter & $\mathcal{U}(0,1)$ & $0.48\pm0.03$ \\
$i$ & Orbital inclination (deg) & \nodata & $86.9\pm0.2$ \\
$R_p/R_\star$ & Radius ratio & $\mathcal{U}(0,1)$ & $0.058\pm0.001$ \\
$K$ & RV semiamplitude (m s$^{-1}$) & $\mathcal{U}(0,1000)$ & $11.2\pm2.3$ \\
$e$ & Eccentricity & \nodata & 0 (fixed, $<0.2$ at $2\sigma$)\\
$a/R_\star$ & Scaled semimajor axis & \nodata & $9.0\pm0.1$ \\
$R_p$ & Planet radius ($R_\oplus$)& \nodata & $9.50\pm0.14$ \\
$a$ & Semimajor axis (au)& \nodata & $0.063\pm0.001$ \\
$M_p$ & Planet mass ($M_\oplus$)& \nodata & $36.3\pm7.3$ \\
$\rho_p$ & Planet density (g cm$^{-3}$)& \nodata & $0.23\pm0.05$ \\
\hline
$q_1^{\rm KPF}$ & KPF linear limb darkening parameter & $\mathcal{U}(0,1)$ & $0.78_{-0.24}^{+0.15}$ \\
$q_2^{\rm KPF}$ & KPF quadratic limb darkening parameter & $\mathcal{U}(0,1)$ & $0.66_{-0.34}^{+0.23}$ \\
$\gamma_{\rm KPF}$ & KPF RV offset (m s$^{-1}$)& $\mathcal{U}(-18500,-18000)$ & $-18226.7\pm2.4$ \\
$\sigma_{\rm KPF}$ & KPF RV jitter (m s$^{-1}$)& $\mathcal{LU}(10^{-3},100)$ & $0.05_{-0.05}^{+0.77}$ \\
$q_1^{\rm NEID}$ & NEID linear limb darkening parameter & $\mathcal{U}(0,1)$ & $0.43_{-0.30}^{+0.37}$ \\
$q_2^{\rm NEID}$ & NEID quadratic limb darkening parameter & $\mathcal{U}(0,1)$ & $0.45_{-0.31}^{+0.36}$ \\
$\gamma_{\rm NEID}$ & NEID RV offset (m s$^{-1}$)& $\mathcal{U}(-250,250)$ & $6.6\pm4.3$ \\
$\sigma_{\rm NEID}$ & NEID RV jitter (m s$^{-1}$)& $\mathcal{LU}(10^{-3},100)$ & $0.20_{-0.19}^{+5.94}$ \\
$\beta$ & Intrinsic stellar line width (km s$^{-1}$) & $\mathcal{G}(6.4,2.0)$ & $6.4\pm2.0$ \\
$\gamma_{\rm SOPHIE}$ & SOPHIE RV offset (m s$^{-1}$)& $\mathcal{U}(-250,250)$ & $0.7\pm1.8$\\
$\sigma_{\rm SOPHIE}$ & SOPHIE RV jitter (m s$^{-1}$)& $\mathcal{LU}(10^{-3},100)$ & $0.1_{-0.1}^{+2.2}$ \\
$\gamma_{\rm HIRES}$ & HIRES RV offset (m s$^{-1}$)& $\mathcal{U}(-250,250)$ & $-7.6\pm6.1$ \\
$\sigma_{\rm HIRES}$ & HIRES RV jitter (m s$^{-1}$)& $\mathcal{LU}(10^{-3},100)$ & $15.1_{-4.1}^{+6.5}$ \\
\hline
$q_1^{\rm TESS}$ & TESS linear limb darkening parameter & $\mathcal{U}(0,1)$ & $0.22_{-0.09}^{+0.18}$ \\
$q_2^{\rm TESS}$ & TESS quadratic limb darkening parameter & $\mathcal{U}(0,1)$ & $0.50_{-0.31}^{+0.32}$ \\
$\sigma_{\rm TESS}^{\rm 30-min}$ & TESS 30-minute cadence photometric jitter (ppm) & $\mathcal{LU}(1,5\times10^7)$ & $9_{-7}^{+36}$ \\
$\sigma_{\rm TESS}^{\rm 2-min}$ & TESS 2-minute cadence photometric jitter (ppm) & $\mathcal{LU}(1,5\times10^7)$ & $10_{-8}^{+43}$ \\
\enddata
\tablecomments{$\mathcal{U}(a,b)$ denotes a uniform prior with a start value $a$ and end value $b$. $\mathcal{G}(\mu,\sigma)$ denotes a normal prior with mean $\mu$, and standard deviation $\sigma$. $\mathcal{LU}(a,b)$ denotes a log-uniform prior with a start value $a$ and end value $b$.}
\end{deluxetable*}

\section{Discussion}\label{sec:discussion}

\subsection{WASP-195 in the Context of Stellar Obliquities}

\citet{Albrecht2021} identified a possible bimodality in the stellar obliquity distribution. Their sample of true obliquities showed seemingly significant clustering around both $\psi\sim0^\circ$ and $\psi\sim90^{\circ}$ \citep[see also][]{Bourrier2023,Attia2023}. Subsequent studies found indications that this preference for polar orbits may be particularly prominent among Neptunes \citep[e.g.,][]{Espinoza-Retamal2024,Knudstrup2024}. These findings have motivated systematic efforts to expand the relatively small sample of stellar-obliquity measurements for Neptune hosts, including the ATREIDES survey \citep{Bourrier2025}, the KPF-SLOPE survey \citep{Handley2026}, and our own POSEIDON survey \citep{Espinoza-Retamal2026,Espinoza-Retamal2026b}.

Additional measurements have weakened the evidence for bimodality. \citet{Espinoza-Retamal2026} found that the available stellar obliquities for Neptune hosts are consistent with a dominant population of well-aligned systems, together with a smaller population showing nearly random obliquities, with no significant clustering at $\sim90^{\circ}$. Interestingly, this distribution resembles that observed for more massive planets \citep[e.g.,][]{Dong2023,Siegel2023,Rossi2025}, suggesting that the populations of transiting Neptunes and Jupiters may share similar dynamical origins \citep[see also][]{WangWang2026}. This possibility is further supported by their similar host-star metallicity distributions \citep[e.g.,][]{Winn2017,Dong2018,Vissapragada2025}, by the apparent enhancement in the occurrence of both populations at orbital periods of a few days \citep[e.g.,][]{Udry2003,Santerne2016,Castro-Gonzalez2024}, and by their similar stellar multiplicity statistics \citep{Eeles-Nolle2025}.

\begin{figure}
    \centering
    \includegraphics[width=\linewidth]{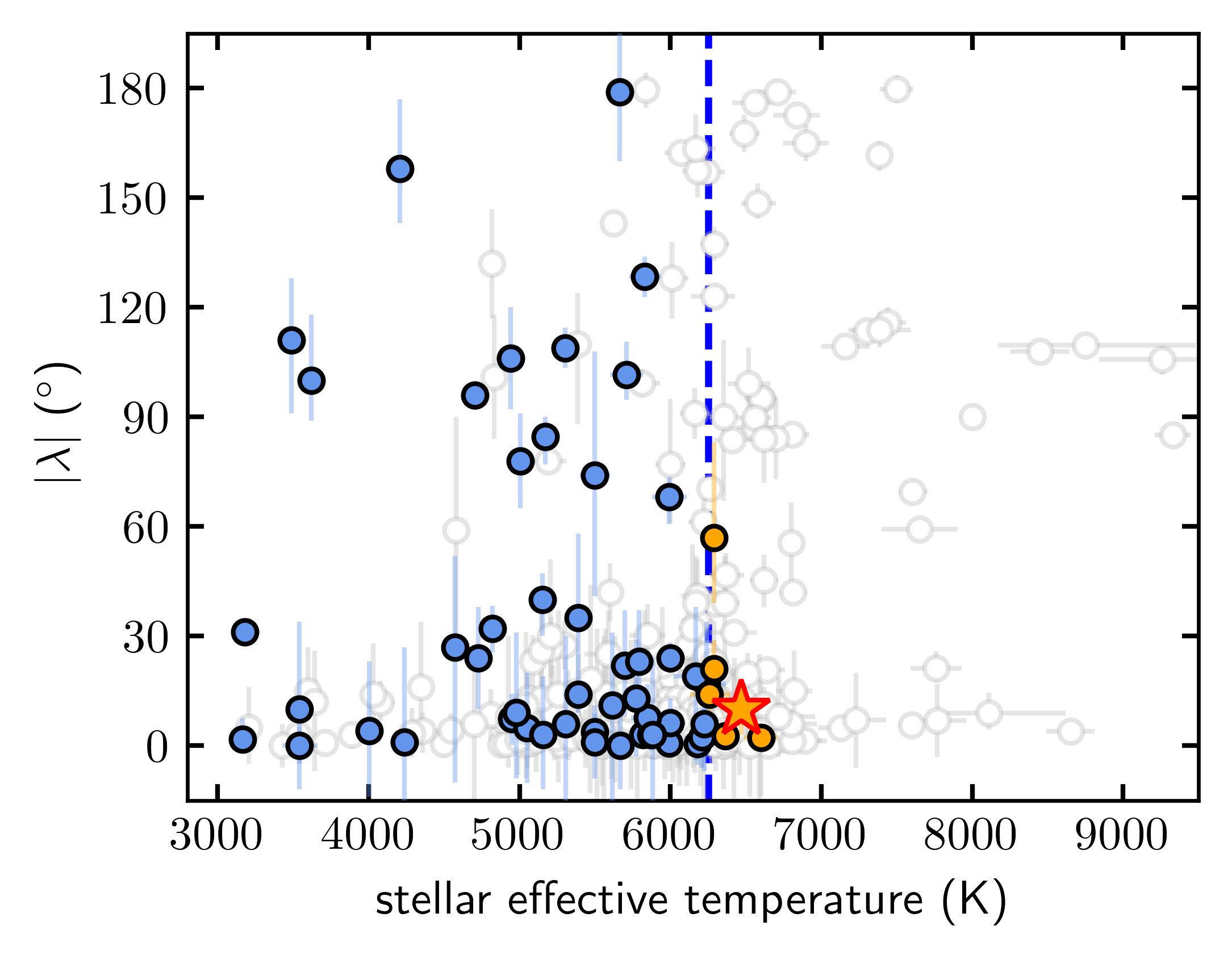}
    \caption{Sky-projected stellar obliquity of Neptune ($2\leq R_p/R_\oplus\leq 6$ or $10\leq M_p/M_\oplus\leq 50$) and Jupiter (not Neptune with $R_p/R_\oplus > 6$) hosts as a function of the stellar effective temperature. The vertical dashed blue line demarks the Kraft break at $T_{\rm eff}\approx6250$~K \citep{Kraft1967}. Neptune hosts below this break are shown in blue, while those above the break are shown in orange. WASP-195 is highlighted as a star with a red edge. Jupiter hosts are shown in grey in the background. Literature data were compiled from SOCat \citep{socat} as of October 5, 2026. Systems flagged by \citet{Albrecht2022} as controversial obliquity measurements, as well as systems with obliquity uncertainties larger than $40^{\circ}$, were excluded from the sample. WASP-195 is one of the few Neptune hosts above the Kraft break with a measured obliquity.}
    \label{fig:Obl_vs_Teff}
\end{figure}

However, an important difference between the current Neptune and Jupiter obliquity samples is in the
distribution of effective temperatures of their
host stars \citep{Dugan2025,Espinoza-Retamal2026,Lafarga2026}. Figure~\ref{fig:Obl_vs_Teff} shows the measured stellar obliquities of Neptune and Jupiter hosts as a function of effective temperature. While the Jupiter sample spans a broad range of stellar temperatures, nearly all Neptunes with measured obliquities orbit stars below the Kraft break at $T_{\rm eff}\approx6250$~K \citep{Kraft1967,Wang2026}. This distinction is particularly important because the obliquity distribution of Jupiters is known to depend strongly on stellar effective temperature, with Jupiters orbiting cool stars being predominantly well aligned, whereas those orbiting hotter stars span a much broader range of obliquities \citep[e.g.,][]{Schlaufman2010,Winn2010,Albrecht2012}.

Expanding the sample of Neptune obliquities above the Kraft break is
therefore essential for determining whether the apparent similarity
between the Neptune and Jupiter obliquity distributions persists when comparing planets orbiting stars in the same temperature regime. WASP-195 provides a valuable addition to this sparsely sampled region of parameter space. With an effective temperature of $T_{\rm eff}=6470\pm100$~K and a rotational velocity $v\sin{i_\star}=10.5\pm1.1$~km~s$^{-1}$, it lies above the Kraft break and is one of the few Neptune hosts in this regime with a measured stellar obliquity. The only other Neptunes around hot stars with measured stellar obliquities are HD~106315~c \citep{Zhou2018,Bourrier2023}, HD~148193~b \citep{Knudstrup2024}, HIP~41378~d and f \citep{Grouffal2022,Grouffal2025}, Kepler-25~c \citep{Albrecht2013,Benomar2014,Campante2016,Bourrier2023}, and TOI-480~b \citep{Handley2026}. Most of these systems are consistent with spin-orbit alignment, with HIP~41378 being a notable exception. However, the sample remains too small to determine whether the obliquity distribution of Neptunes around hot stars differs from that observed around cooler stars, or whether it resembles the broad distribution observed for Jupiters around hot stars. As the POSEIDON survey continues, we will further expand the obliquity census into regions of planetary and stellar parameter space that remain poorly sampled. 

\subsection{The Missing Neptunes Around Hot Stars}\label{sec:missing_neptunes}

\begin{figure}
    \centering
    \includegraphics[width=\linewidth]{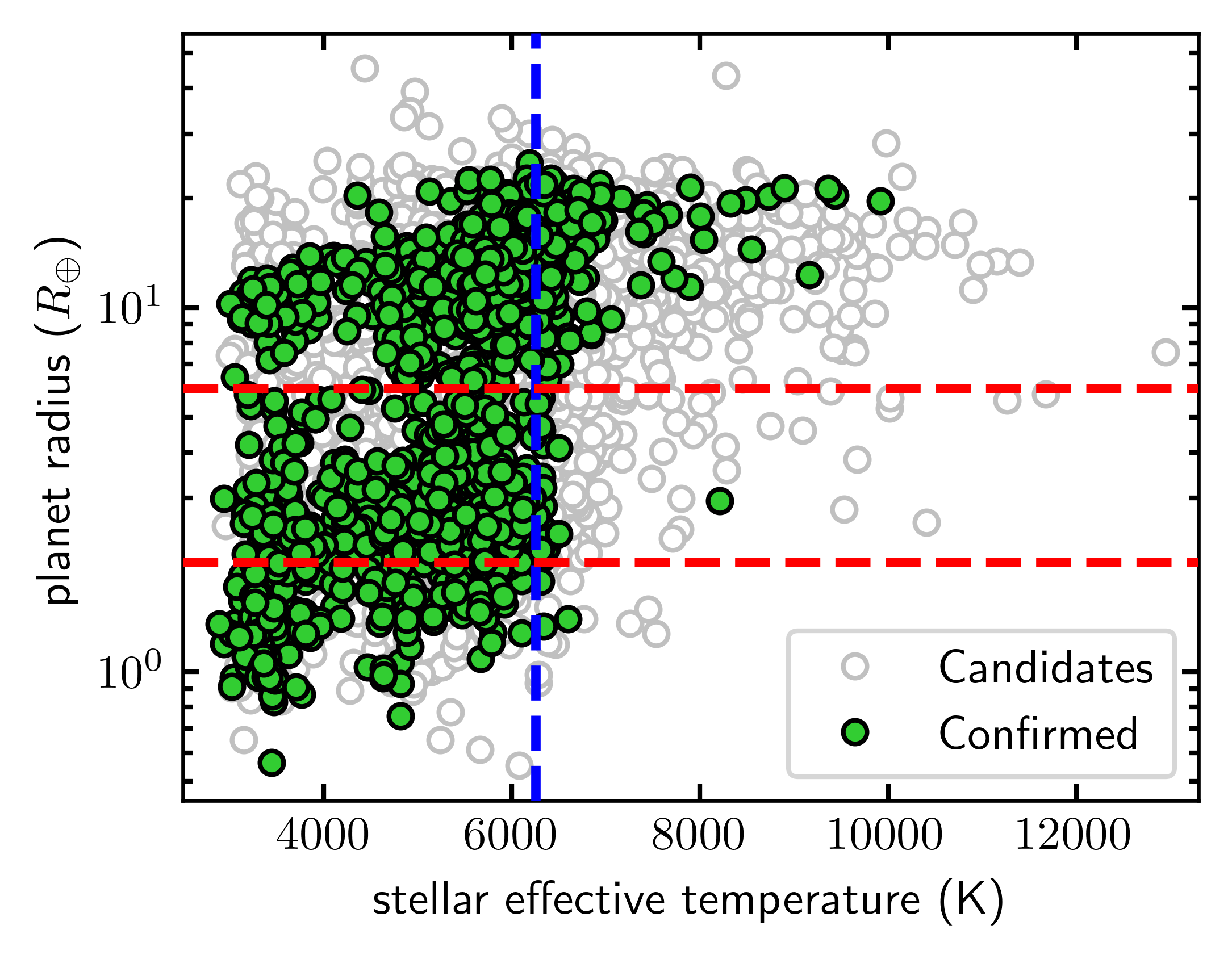}
    \caption{Planet radius versus stellar effective temperature diagram for TESS Objects of Interest. Planet candidates are shown in grey. Confirmed and known planets are shown in green. Known false positives are not considered for the plot. The vertical dashed blue line demarks the Kraft break at $T_{\rm eff}\approx6250$~K \citep{Kraft1967}. The horizontal dashed red lines correspond to planet radii of 2 and 6~$R_{\oplus}$, which we adopt as the Neptune-size regime. Data were obtained from ExoFOP as of May 2026.}
    \label{fig:TOIs}
\end{figure}

The scarcity of obliquity measurements for Neptune hosts above the Kraft break may be due in part to the observational challenges associated with measuring the RM effect in these systems \citep[see, e.g.,][]{Hartman2015,Temple2018,Dugan2025}. However, at present, the fundamental limitation is that there are relatively few confirmed Neptunes around hot stars available for spin-orbit measurements in the first place.
Figure~\ref{fig:TOIs} shows the distribution of TESS Objects of Interest as a function of planetary radius and stellar effective temperature, separated into planet candidates and confirmed planets. While numerous Neptune-sized candidates have been identified around stars above the Kraft break, comparatively few planets have been confirmed in this region of parameter space. Therefore, the scarcity of obliquity measurements for Neptunes around hot stars is at least partly inherited from the small population of confirmed planets available for follow-up.

Confirming the presence of Neptune-sized planets around hot stars is particularly challenging with RV observations. Their relatively low masses produce small orbital RV signals, while hot stars typically rotate more rapidly and have broader spectral lines, degrading the precision of RV observations. Broad spectral lines can also complicate the identification of additional spectral components from unresolved stellar companions, making it more difficult to rule out some false-positive scenarios \citep[see, e.g.,][]{Mandushev2005,Winn2025}. These effects make it particularly challenging to establish the planetary nature and measure the masses of Neptune candidates around hot stars. Nevertheless, the existence of systems such as HD~56414~b, a warm Neptune orbiting an A-type star \citep{Giacalone2022}, and Kepler-462~b and c \citep{Ahlers2015}, together with the population of candidates shown in Figure~\ref{fig:TOIs}, demonstrates that Neptune-sized planets can form and survive around hot stars.

At the same time, the scarcity of these planets does not appear to be purely observational. Previous demographic studies have found that the occurrence rate of short-period small planets decreases toward higher stellar masses and effective temperatures \citep[e.g.,][]{Mulders2015,Hardegree-Ullman2025}. In particular, \citet{Giacalone2025} searched more than 20,000 A-type stars observed by TESS and found no reliable planets smaller than $8\,R_\oplus$ with orbital periods shorter than 10 days, placing upper limits on the occurrence of close-in sub-Neptunes and sub-Saturns that are 3--6 times lower than estimates for Sun-like stars. Thus, although observational selection effects contribute to the small number of confirmed Neptunes around hot stars, there is also evidence that close-in planets in this size regime are intrinsically less common around hotter and more massive stars.

One possible explanation for this trend is that the conditions in the inner regions of protoplanetary disks depend strongly on stellar mass and luminosity. The characteristic radii associated with magnetospheric disk truncation and dust sublimation move outward around more massive and luminous stars, potentially preventing low-mass planets from migrating to the shortest orbital periods \citep[e.g.,][]{Mulders2015,Giacalone2025}. Recent models of planet migration in the inner disks of young stars support this picture, predicting that lower-mass planets can stall near the dust-sublimation radius or the inner boundary of the
low-ionization region (``dead zone''), at larger orbital separations around more massive stars \citep{CevallosSoto2026}. This may help explain why the few known Neptune-sized planets orbiting hot stars tend to have relatively long orbital periods, while systems such as WASP-195~b appear to be rare.

Interestingly, the observational biases affecting planetary confirmation and obliquity measurements do not act in the same direction. Rapid stellar rotation degrades the precision of RV measurements, making low-mass planets more difficult to confirm. At the same time, the amplitude of the RM effect scales with $v\sin{i_\star}$ \citep[e.g.,][]{Triaud2018}, such that rapid stellar rotation increases the amplitude of the spectroscopic transit signal. As noted by \citet{Gaudi2007}, the RM effect can therefore provide an additional avenue for establishing the planetary nature of transiting candidates in cases where rapid stellar rotation prevents a precise detection of the orbital RV signal. This approach may be particularly valuable for Neptune candidates around hot, rapidly rotating stars, while simultaneously providing information about their spin--orbit architectures.


\subsection{Search for Additional Companions}

One fundamental question that arises when
trying to understand the formation and dynamical evolution of a system is whether there are additional massive companions. \citet{Schanche2025} obtained high-resolution speckle imaging of WASP-195 and detected no nearby stellar companions. The observations rule out companions brighter than $\sim5$--$6$ mag below that of WASP-195 at angular separations of $0.5$--$1.2^{\prime\prime}$, corresponding to projected separations of $\sim250$--$590$ au. This lack of nearby stellar companions is also supported by the Gaia DR3 Renormalized Unit Weight Error \citep[RUWE;][]{GaiaDR3} value of $\sim0.85$, which suggests that any unresolved companions do not cause detectable astrometric perturbations. Regarding wider separations, WASP-195 is not listed in the wide binary catalog of \citet{El-Badry2021}, which is based on Gaia Early DR3 proper motions and parallaxes \citep{GaiaEDR3}, indicating the absence of co-moving stellar companions out to separations of $\sim1$~pc. The available data support a single-star configuration for WASP-195.

\begin{figure}
    \centering
    \includegraphics[width=\linewidth]{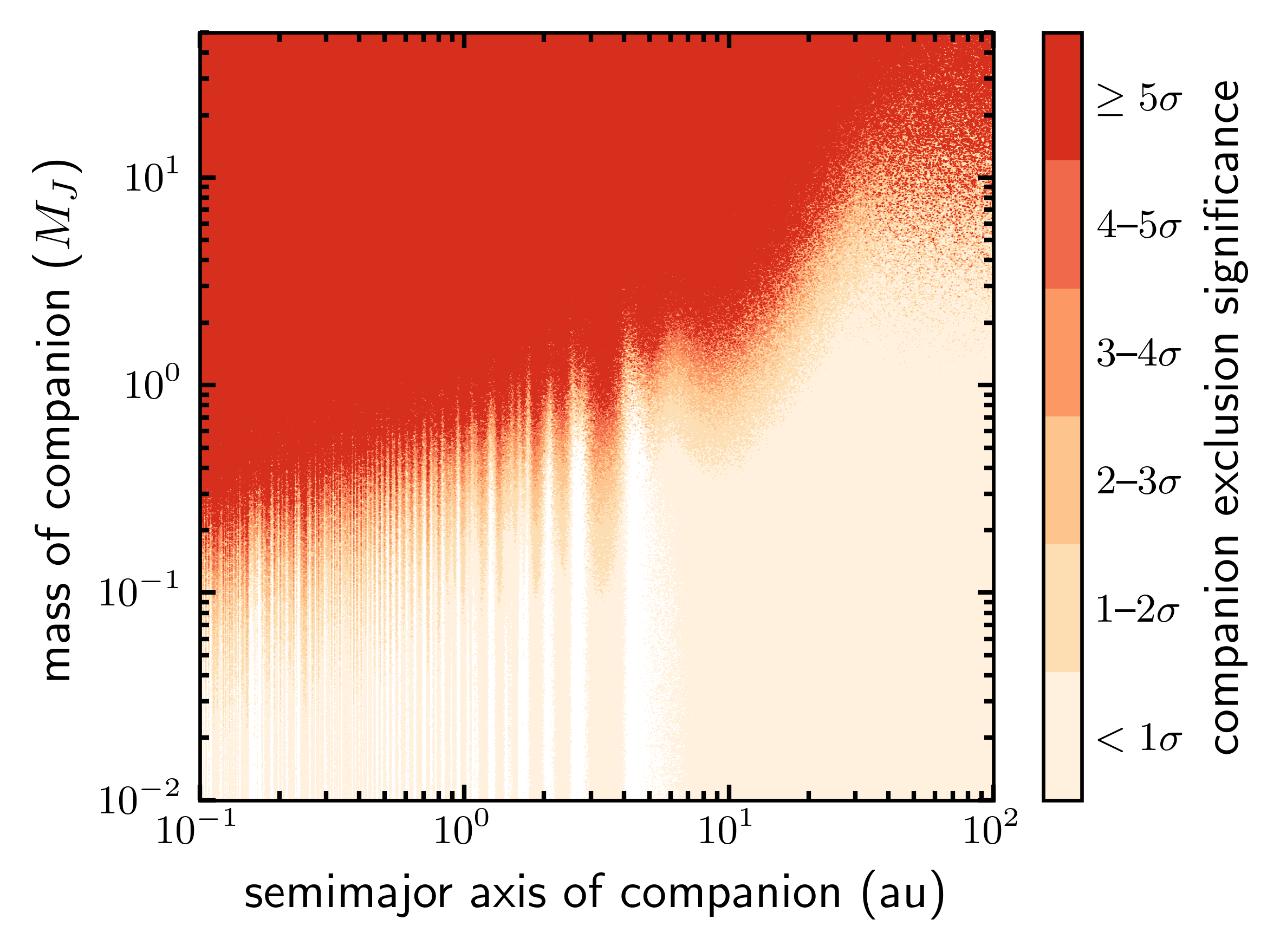}
    \caption{Mass versus semimajor axis diagram of companions that can be excluded, given the lack of additional signals in the SOPHIE RV residuals. Different colors represent different confidence levels. Based on the 9.3 years of SOPHIE RV observations, we can rule out Jupiter-mass companions to WASP-195 within $\sim3$~au at a $5\sigma$ confidence.}
    \label{fig:companions}
\end{figure}

As for additional planetary companions, there is no evidence for TTVs (see Section~\ref{sec:phot}), and a box least squares algorithm did not reveal any significant additional transit signals. Similarly, the SOPHIE RVs do not show any long-term trends or additional periodic signals. Following the approach of \citet{Espinoza-Retamal2024}, we performed a population synthesis analysis to quantify the types of planetary companions that can be ruled out by our data. In brief, we generated a synthetic population of companions with masses between 0.01 and 50~$M_J$ and semimajor axes between 0.01 and 100~au, assuming isotropic orbital orientations, eccentricities drawn from the \citet{Kipping2013_ecc} distribution, and arguments of periastron drawn from a uniform distribution. The RV signals induced by these synthetic companions were compared to the observed SOPHIE RV residuals using a $\chi^2$ metric to determine which companions could be excluded at different confidence levels. Figure~\ref{fig:companions} shows the regions of the mass versus semimajor axis plane that can be excluded at different confidence levels. With a baseline of 9.3 years of SOPHIE observations, the RV data rule out the presence of Jupiter-mass companions to WASP-195 within $\sim3$~au at a $5\sigma$ confidence.

\subsection{The Dynamical History of WASP-195 b}

The proposed mechanisms to explain the existence of short-period exoplanets such as WASP-195~b generally fall into two broad groups: disk-driven migration and high-eccentricity or tidally-driven migration \citep[see, e.g.,][]{Dawson2018}. In the specific case of WASP-195~b, its short-period, aligned, and nearly circular orbit ($e<0.2$ at $2\sigma$ confidence) is broadly consistent with disk-driven evolution within a primordially aligned protoplanetary disk. Therefore, both in-situ formation \citep[e.g.,][]{Batygin2016,Boley2016} and inward migration driven by nebular tides \citep[e.g.,][]{Goldreich1980,Lin1986,Ward1997} are viable pathways for producing the observed system architecture. In particular, the low obliquity of WASP-195~b is consistent with the picture discussed in Section~\ref{sec:missing_neptunes}, in which the rare Neptune-mass planets that reach short orbital periods around hot stars through disk migration are expected to preserve their primordial spin-orbit alignment.

High-eccentricity tidal migration could, in principle, provide an alternative explanation for the observed architecture of the system. Recent theoretical work has studied the role of high-eccentricity migration in shaping the population of short-period Neptunes, while emphasizing the importance of tidal-structural evolution during this process \citep[e.g.,][]{Castro-Gonzalez2026,Hallatt2026b,Zanazzi2026}. Following \citet{Goldreich66} and \citet{Hut81}, and adopting a modified tidal quality factor \citep[e.g.,][]{Ogilvie07} of $Q^\prime = 10^4-10^5$, we estimate a tidal circularization timescale of $\sim40-400$~Myr. This timescale is comfortably shorter than the estimated age of the system of 2.1~Gyr, indicating that an initially eccentric orbit probably would have been circularized through tidal dissipation. Although high-eccentricity migration mechanisms are typically associated with large stellar obliquities, a significant fraction of systems are expected to end up with low obliquities. For example, \citet{Veldhuis2025} found that $\sim20\%$ of systems undergoing von Zeipel--Lidov--Kozai oscillations end with obliquities $\lesssim40^\circ$ \citep[see also the distributions presented by][]{Beauge2012,Anderson2016,Teyssandier2019}. Therefore, the low projected obliquity of WASP-195 does not by itself rule out a high-eccentricity migration origin.

An alternative high-eccentricity migration pathway that naturally produces low stellar obliquities is coplanar high-eccentricity migration \citep{Petrovich2015,Bhaskar2026}, which can preserve low obliquities throughout the dynamical evolution. This scenario requires a massive, eccentric, and nearly coplanar long-period companion to excite the planet's eccentricity and trigger tidal migration. Our RV observations exclude Jupiter-mass companions within $\sim3$ au, although such a planet could remain undetected at larger orbital separations. Alternatively, the system could have undergone tidal realignment after migration. However, tidal realignment timescales for Neptune-mass planets are expected to be orders of magnitude longer than the age of the universe \citep[e.g.,][]{Zahn1977,Albrecht2012}, especially for hot stars near or above the Kraft break, such as WASP-195, making tidal realignment unlikely in this system. A distinct evolutionary pathway involves Roche-lobe overflow from a more massive progenitor \citep[e.g.,][]{Valsecchi2015,Hallatt2025}, during which angular momentum transfer may drive the system toward spin-orbit alignment \citep{Hallatt2026}. Overall, the observed architecture of WASP-195~b is naturally consistent with disk-driven migration, although alternative pathways such as coplanar high-eccentricity migration and Roche-lobe overflow cannot be excluded.

\section{Summary and Conclusions}\label{sec:conclusion}

In this third paper of the POSEIDON survey, we presented new spectroscopic transit observations of WASP-195~b obtained with KPF and NEID, together with new HIRES RVs. By jointly modeling all these new and archival observations, we measured the projected stellar obliquity of the system, refined its orbital and planetary properties, and searched for additional companions.

We find that WASP-195~b orbits its host star on a well-aligned orbit with a projected stellar obliquity $\lambda=-10\pm7^\circ$, making it one of the few Neptunes with a host star hotter than the Kraft break and an empirical obliquity constraint. The available RV observations show no evidence for additional companions, allowing us to rule out Jupiter-mass planets within $\sim3$ au at the $5\sigma$ confidence level. The aligned and nearly circular architecture of the system is naturally explained by disk-driven migration, although alternative scenarios such as coplanar high-eccentricity migration and Roche-lobe overflow cannot yet be excluded.

Beyond characterizing this system, our analysis suggests that the scarcity of Neptune systems with measured stellar obliquities above the Kraft break reflects a combination of the lower intrinsic occurrence of short-period Neptunes around hot stars and the observational challenge of confirming small planets orbiting hot, rapidly rotating stars.
Observations of the RM effect may be helpful for both confirming the planetary nature of transiting candidates and constraining their stellar obliquities.

\begin{acknowledgments}

\added{We thank the anonymous referee for their careful reading of the manuscript and constructive comments.}

This work was supported by a NASA-Keck PI Data Award, administered by the NASA Exoplanet Science Institute. Data presented herein were obtained at the W.\ M.\ Keck Observatory from telescope time allocated to the National Aeronautics and Space Administration through the agency's scientific partnership with the California Institute of Technology and the University of California. The Observatory was made possible by the generous financial support of the W.\ M.\ Keck Foundation. The authors wish to recognize and acknowledge the very significant cultural role and reverence that the summit of Maunakea has always had within the indigenous Hawaiian community. We are most fortunate to have the opportunity to conduct observations from this mountain.

This work is based on observations taken with the NEID instrument on the WIYN 3.5 m telescope at Kitt Peak National Observatory. We thank the NEID Queue Observers and WIYN Observing Associates for their skillful execution of our NEID observations. The authors are honored to be permitted to conduct astronomical research on I'oligam Du'ag (Kitt Peak), a mountain with particular significance to the Tohono O'odham. Kitt Peak is a facility of NSF's NOIRLab, managed by the Association of Universities for Research in Astronomy (AURA). The WIYN telescope is a joint facility of NOIRLab, Indiana University, the University of Wisconsin-Madison, Pennsylvania State University, Purdue University, and Princeton University. NEID was funded by the NASA-NSF Exoplanet Observational Research (NN-EXPLORE) partnership and built by Pennsylvania State University. The NEID archive is operated by the NASA Exoplanet Science Institute at the California Institute of Technology. NN-EXPLORE is managed by the Jet Propulsion Laboratory, California Institute of Technology under contract with NASA. 

This work includes data collected with the TESS mission, obtained from the MAST data archive at the Space Telescope Science Institute (STScI). Funding for the TESS mission is provided by the NASA Explorer Program. STScI is operated by AURA, Inc., under NASA contract NAS 5–26555.

This research was carried out, in part, at the Jet Propulsion Laboratory and the California Institute of Technology under a contract with NASA and funded through the President’s and Director’s Research \& Development Fund Program.

The authors are pleased to acknowledge that the work reported on in this paper was substantially performed using Princeton University’s Research Computing resources.

J.I.E.-R.\ acknowledges support from the Heising-Simons Foundation 51 Pegasi b Fellowship.

R.B.\ acknowledges support from FONDECYT Project 1241963.

A.J.\ acknowledges support from Fondecyt project 1251439.

X.-Y.W.\ acknowledges support from the Sullivan Prize Fellowship.

N.S.\ acknowledges support from the Yale Center for Astronomy \& Astrophysics Prize Postdoctoral Fellowship.

S.G.\ is supported by an NSF Astronomy and Astrophysics Postdoctoral Fellowship under award AST2303922.

\facilities{Keck:I (KPF, HIRES), WIYN (NEID), OHP:1.93m (SOPHIE), TESS}

\software{
\texttt{astropy} \citep{astropy,astropy2,astropy3},
\texttt{batman} \citep{batman},
\texttt{celerite} \citep{celerite},
\texttt{dynesty} \citep{dynesty2},
\texttt{ironman} \citep{Espinoza-Retamal2023b,Espinoza-Retamal2024},
\texttt{juliet} \citep{juliet},
\texttt{lightkurve} \citep{lightkurve},
\texttt{matplotlib} \citep{matplotlib},
\texttt{numpy} \citep{numpy},
\texttt{radvel} \citep{Fulton2018},
\texttt{rmfit} \citep{Stefansson2022},
\texttt{scipy} \citep{scipy},
\texttt{serval} \citep{Zechmeister2018},
\texttt{zaspe} \citep{zaspe}.
}

\end{acknowledgments}

\bibliography{sample7}{}

\begin{thebibliography}{}
\expandafter\ifx\csname natexlab\endcsname\relax\def\natexlab#1{#1}\fi
\providecommand{\url}[1]{\href{#1}{#1}}
\providecommand{\dodoi}[1]{doi:~\href{http://doi.org/#1}{\nolinkurl{#1}}}
\providecommand{\doeprint}[1]{\href{http://ascl.net/#1}{\nolinkurl{http://ascl.net/#1}}}
\providecommand{\doarXiv}[1]{\href{https://arxiv.org/abs/#1}{\nolinkurl{https://arxiv.org/abs/#1}}}

\bibitem[{J.~P. {Ahlers} {et~al.}(2015){Ahlers}, {Barnes}, \& {Barnes}}]{Ahlers2015}
{Ahlers}, J.~P., {Barnes}, J.~W., \& {Barnes}, R. 2015, \bibinfo{title}{{Spin-Orbit Misalignment of Two-Planet-System KOI-89 Via Gravity Darkening},} \apj, 814, 67, \dodoi{10.1088/0004-637X/814/1/67}

\bibitem[{S. {Albrecht} {et~al.}(2013){Albrecht}, {Winn}, {Marcy}, {Howard}, {Isaacson}, \& {Johnson}}]{Albrecht2013}
{Albrecht}, S., {Winn}, J.~N., {Marcy}, G.~W., {et~al.} 2013, \bibinfo{title}{{Low Stellar Obliquities in Compact Multiplanet Systems},} \apj, 771, 11, \dodoi{10.1088/0004-637X/771/1/11}

\bibitem[{S. {Albrecht} {et~al.}(2012){Albrecht}, {Winn}, {Johnson}, {Howard}, {Marcy}, {Butler}, {Arriagada}, {Crane}, {Shectman}, {Thompson}, {Hirano}, {Bakos}, \& {Hartman}}]{Albrecht2012}
{Albrecht}, S., {Winn}, J.~N., {Johnson}, J.~A., {et~al.} 2012, \bibinfo{title}{{Obliquities of Hot Jupiter Host Stars: Evidence for Tidal Interactions and Primordial Misalignments},} \apj, 757, 18, \dodoi{10.1088/0004-637X/757/1/18}

\bibitem[{S.~H. {Albrecht} {et~al.}(2022){Albrecht}, {Dawson}, \& {Winn}}]{Albrecht2022}
{Albrecht}, S.~H., {Dawson}, R.~I., \& {Winn}, J.~N. 2022, \bibinfo{title}{{Stellar Obliquities in Exoplanetary Systems},} \pasp, 134, 082001, \dodoi{10.1088/1538-3873/ac6c09}

\bibitem[{S.~H. {Albrecht} {et~al.}(2021){Albrecht}, {Marcussen}, {Winn}, {Dawson}, \& {Knudstrup}}]{Albrecht2021}
{Albrecht}, S.~H., {Marcussen}, M.~L., {Winn}, J.~N., {Dawson}, R.~I., \& {Knudstrup}, E. 2021, \bibinfo{title}{{A Preponderance of Perpendicular Planets},} \apjl, 916, L1, \dodoi{10.3847/2041-8213/ac0f03}

\bibitem[{K.~R. {Anderson} {et~al.}(2016){Anderson}, {Storch}, \& {Lai}}]{Anderson2016}
{Anderson}, K.~R., {Storch}, N.~I., \& {Lai}, D. 2016, \bibinfo{title}{{Formation and stellar spin-orbit misalignment of hot Jupiters from Lidov-Kozai oscillations in stellar binaries},} \mnras, 456, 3671, \dodoi{10.1093/mnras/stv2906}

\bibitem[{S.~M. {Andrews}(2020){Andrews}}]{Andrews2020}
{Andrews}, S.~M. 2020, \bibinfo{title}{{Observations of Protoplanetary Disk Structures},} \araa, 58, 483, \dodoi{10.1146/annurev-astro-031220-010302}

\bibitem[{ {Astropy Collaboration} {et~al.}(2013){Astropy Collaboration}, {Robitaille}, {Tollerud}, {Greenfield}, {Droettboom}, {Bray}, {Aldcroft}, {Davis}, {Ginsburg}, {Price-Whelan}, {Kerzendorf}, {Conley}, {Crighton}, {Barbary}, {Muna}, {Ferguson}, {Grollier}, {Parikh}, {Nair}, {Unther}, {Deil}, {Woillez}, {Conseil}, {Kramer}, {Turner}, {Singer}, {Fox}, {Weaver}, {Zabalza}, {Edwards}, {Azalee Bostroem}, {Burke}, {Casey}, {Crawford}, {Dencheva}, {Ely}, {Jenness}, {Labrie}, {Lim}, {Pierfederici}, {Pontzen}, {Ptak}, {Refsdal}, {Servillat}, \& {Streicher}}]{astropy}
{Astropy Collaboration}, {Robitaille}, T.~P., {Tollerud}, E.~J., {et~al.} 2013, \bibinfo{title}{{Astropy: A community Python package for astronomy},} \aap, 558, A33, \dodoi{10.1051/0004-6361/201322068}

\bibitem[{ {Astropy Collaboration} {et~al.}(2018){Astropy Collaboration}, {Price-Whelan}, {Sip{\H{o}}cz}, {G{\"u}nther}, {Lim}, {Crawford}, {Conseil}, {Shupe}, {Craig}, {Dencheva}, {Ginsburg}, {VanderPlas}, {Bradley}, {P{\'e}rez-Su{\'a}rez}, {de Val-Borro}, {Aldcroft}, {Cruz}, {Robitaille}, {Tollerud}, {Ardelean}, {Babej}, {Bach}, {Bachetti}, {Bakanov}, {Bamford}, {Barentsen}, {Barmby}, {Baumbach}, {Berry}, {Biscani}, {Boquien}, {Bostroem}, {Bouma}, {Brammer}, {Bray}, {Breytenbach}, {Buddelmeijer}, {Burke}, {Calderone}, {Cano Rodr{\'\i}guez}, {Cara}, {Cardoso}, {Cheedella}, {Copin}, {Corrales}, {Crichton}, {D'Avella}, {Deil}, {Depagne}, {Dietrich}, {Donath}, {Droettboom}, {Earl}, {Erben}, {Fabbro}, {Ferreira}, {Finethy}, {Fox}, {Garrison}, {Gibbons}, {Goldstein}, {Gommers}, {Greco}, {Greenfield}, {Groener}, {Grollier}, {Hagen}, {Hirst}, {Homeier}, {Horton}, {Hosseinzadeh}, {Hu}, {Hunkeler}, {Ivezi{\'c}}, {Jain}, {Jenness}, {Kanarek}, {Kendrew}, {Kern}, {Kerzendorf}, {Khvalko}, {King}, {Kirkby}, {Kulkarni},
  {Kumar}, {Lee}, {Lenz}, {Littlefair}, {Ma}, {Macleod}, {Mastropietro}, {McCully}, {Montagnac}, {Morris}, {Mueller}, {Mumford}, {Muna}, {Murphy}, {Nelson}, {Nguyen}, {Ninan}, {N{\"o}the}, {Ogaz}, {Oh}, {Parejko}, {Parley}, {Pascual}, {Patil}, {Patil}, {Plunkett}, {Prochaska}, {Rastogi}, {Reddy Janga}, {Sabater}, {Sakurikar}, {Seifert}, {Sherbert}, {Sherwood-Taylor}, {Shih}, {Sick}, {Silbiger}, {Singanamalla}, {Singer}, {Sladen}, {Sooley}, {Sornarajah}, {Streicher}, {Teuben}, {Thomas}, {Tremblay}, {Turner}, {Terr{\'o}n}, {van Kerkwijk}, {de la Vega}, {Watkins}, {Weaver}, {Whitmore}, {Woillez}, {Zabalza}, \& {Astropy Contributors}}]{astropy2}
{Astropy Collaboration}, {Price-Whelan}, A.~M., {Sip{\H{o}}cz}, B.~M., {et~al.} 2018, \bibinfo{title}{{The Astropy Project: Building an Open-science Project and Status of the v2.0 Core Package},} \aj, 156, 123, \dodoi{10.3847/1538-3881/aabc4f}

\bibitem[{ {Astropy Collaboration} {et~al.}(2022){Astropy Collaboration}, {Price-Whelan}, {Lim}, {Earl}, {Starkman}, {Bradley}, {Shupe}, {Patil}, {Corrales}, {Brasseur}, {N{\"o}the}, {Donath}, {Tollerud}, {Morris}, {Ginsburg}, {Vaher}, {Weaver}, {Tocknell}, {Jamieson}, {van Kerkwijk}, {Robitaille}, {Merry}, {Bachetti}, {G{\"u}nther}, {Aldcroft}, {Alvarado-Montes}, {Archibald}, {B{\'o}di}, {Bapat}, {Barentsen}, {Baz{\'a}n}, {Biswas}, {Boquien}, {Burke}, {Cara}, {Cara}, {Conroy}, {Conseil}, {Craig}, {Cross}, {Cruz}, {D'Eugenio}, {Dencheva}, {Devillepoix}, {Dietrich}, {Eigenbrot}, {Erben}, {Ferreira}, {Foreman-Mackey}, {Fox}, {Freij}, {Garg}, {Geda}, {Glattly}, {Gondhalekar}, {Gordon}, {Grant}, {Greenfield}, {Groener}, {Guest}, {Gurovich}, {Handberg}, {Hart}, {Hatfield-Dodds}, {Homeier}, {Hosseinzadeh}, {Jenness}, {Jones}, {Joseph}, {Kalmbach}, {Karamehmetoglu}, {Ka{\l}uszy{\'n}ski}, {Kelley}, {Kern}, {Kerzendorf}, {Koch}, {Kulumani}, {Lee}, {Ly}, {Ma}, {MacBride}, {Maljaars}, {Muna}, {Murphy}, {Norman},
  {O'Steen}, {Oman}, {Pacifici}, {Pascual}, {Pascual-Granado}, {Patil}, {Perren}, {Pickering}, {Rastogi}, {Roulston}, {Ryan}, {Rykoff}, {Sabater}, {Sakurikar}, {Salgado}, {Sanghi}, {Saunders}, {Savchenko}, {Schwardt}, {Seifert-Eckert}, {Shih}, {Jain}, {Shukla}, {Sick}, {Simpson}, {Singanamalla}, {Singer}, {Singhal}, {Sinha}, {Sip{\H{o}}cz}, {Spitler}, {Stansby}, {Streicher}, {{\v{S}}umak}, {Swinbank}, {Taranu}, {Tewary}, {Tremblay}, {de Val-Borro}, {Van Kooten}, {Vasovi{\'c}}, {Verma}, {de Miranda Cardoso}, {Williams}, {Wilson}, {Winkel}, {Wood-Vasey}, {Xue}, {Yoachim}, {Zhang}, {Zonca}, \& {Astropy Project Contributors}}]{astropy3}
{Astropy Collaboration}, {Price-Whelan}, A.~M., {Lim}, P.~L., {et~al.} 2022, \bibinfo{title}{{The Astropy Project: Sustaining and Growing a Community-oriented Open-source Project and the Latest Major Release (v5.0) of the Core Package},} \apj, 935, 167, \dodoi{10.3847/1538-4357/ac7c74}

\bibitem[{M. {Attia} {et~al.}(2023){Attia}, {Bourrier}, {Delisle}, \& {Eggenberger}}]{Attia2023}
{Attia}, M., {Bourrier}, V., {Delisle}, J.-B., \& {Eggenberger}, P. 2023, \bibinfo{title}{{DREAM: II. The spin{\textendash}orbit angle distribution of close-in exoplanets under the lens of tides},} \aap, 674, A120, \dodoi{10.1051/0004-6361/202245237}

\bibitem[{K. {Batygin} {et~al.}(2016){Batygin}, {Bodenheimer}, \& {Laughlin}}]{Batygin2016}
{Batygin}, K., {Bodenheimer}, P.~H., \& {Laughlin}, G.~P. 2016, \bibinfo{title}{{In Situ Formation and Dynamical Evolution of Hot Jupiter Systems},} \apj, 829, 114, \dodoi{10.3847/0004-637X/829/2/114}

\bibitem[{C. {Beaug{\'e}} \& D. {Nesvorn{\'y}}(2012){Beaug{\'e}} \& {Nesvorn{\'y}}}]{Beauge2012}
{Beaug{\'e}}, C., \& {Nesvorn{\'y}}, D. 2012, \bibinfo{title}{{Multiple-planet Scattering and the Origin of Hot Jupiters},} \apj, 751, 119, \dodoi{10.1088/0004-637X/751/2/119}

\bibitem[{J.~G. {Beck} \& P. {Giles}(2005){Beck} \& {Giles}}]{Beck2005}
{Beck}, J.~G., \& {Giles}, P. 2005, \bibinfo{title}{{Helioseismic Determination of the Solar Rotation Axis},} \apjl, 621, L153, \dodoi{10.1086/429224}

\bibitem[{C. {Bender} {et~al.}(2022){Bender}, {Ninan}, {Terrien}, {Roy}, {Esplin}, {Kaplan}, {Ca{\~n}as}, {Fulton}, {Gupta}, {Halverson}, {Kanodia}, {Laher}, {Lin}, {Salazar Rivera}, {Blake}, {Diddams}, {Gong}, {Hearty}, {Li}, {Logsdon}, {Lubar}, {Mahadevan}, {McElwain}, {Monson}, {Nitroy}, {Rajagopal}, {Ramsey}, {Robertson}, {Schwab}, {Stefansson}, \& {Wright}}]{Bender2022}
{Bender}, C., {Ninan}, J., {Terrien}, R., {et~al.} 2022, \bibinfo{title}{{Overview and Current Status of the NEID Data Reduction Pipeline},} in American Astronomical Society Meeting Abstracts, Vol. 240, American Astronomical Society Meeting \#240, 401.01

\bibitem[{O. {Benomar} {et~al.}(2014){Benomar}, {Masuda}, {Shibahashi}, \& {Suto}}]{Benomar2014}
{Benomar}, O., {Masuda}, K., {Shibahashi}, H., \& {Suto}, Y. 2014, \bibinfo{title}{{Determination of three-dimensional spin-orbit angle with joint analysis of asteroseismology, transit lightcurve, and the Rossiter-McLaughlin effect: Cases of HAT-P-7 and Kepler-25},} \pasj, 66, 94, \dodoi{10.1093/pasj/psu069}

\bibitem[{H. {Bhaskar} {et~al.}(2026){Bhaskar}, {Petrovich}, \& {Mu{\~n}oz}}]{Bhaskar2026}
{Bhaskar}, H., {Petrovich}, C., \& {Mu{\~n}oz}, D.~J. 2026, \bibinfo{title}{{Planet-Planet Secular Migration Predicts a Stellar Obliquity-Period Anti-Correlation},} arXiv e-prints, arXiv:2606.20803, \dodoi{10.48550/arXiv.2606.20803}

\bibitem[{A.~C. {Boley} {et~al.}(2016){Boley}, {Granados Contreras}, \& {Gladman}}]{Boley2016}
{Boley}, A.~C., {Granados Contreras}, A.~P., \& {Gladman}, B. 2016, \bibinfo{title}{{The In Situ Formation of Giant Planets at Short Orbital Periods},} \apjl, 817, L17, \dodoi{10.3847/2041-8205/817/2/L17}

\bibitem[{F. {Bouchy} {et~al.}(2009){Bouchy}, {H{\'e}brard}, {Udry}, {Delfosse}, {Boisse}, {Desort}, {Bonfils}, {Eggenberger}, {Ehrenreich}, {Forveille}, {Lagrange}, {Le Coroller}, {Lovis}, {Moutou}, {Pepe}, {Perrier}, {Pont}, {Queloz}, {Santos}, {S{\'e}gransan}, \& {Vidal-Madjar}}]{Bouchy2009}
{Bouchy}, F., {H{\'e}brard}, G., {Udry}, S., {et~al.} 2009, \bibinfo{title}{{The SOPHIE search for northern extrasolar planets . I. A companion around HD 16760 with mass close to the planet/brown-dwarf transition},} \aap, 505, 853, \dodoi{10.1051/0004-6361/200912427}

\bibitem[{V. {Bourrier} {et~al.}(2023){Bourrier}, {Attia}, {Mallonn}, {Marret}, {Lendl}, {Konig}, {Krenn}, {Cretignier}, {Allart}, {Henry}, {Bryant}, {Leleu}, {Nielsen}, {Hebrard}, {Hara}, {Ehrenreich}, {Seidel}, {dos Santos}, {Lovis}, {Bayliss}, {Cegla}, {Dumusque}, {Boisse}, {Boucher}, {Bouchy}, {Pepe}, {Lavie}, {Rey Cerda}, {S{\'e}gransan}, {Udry}, \& {Vrignaud}}]{Bourrier2023}
{Bourrier}, V., {Attia}, M., {Mallonn}, M., {et~al.} 2023, \bibinfo{title}{{DREAM: I. Orbital architecture orrery},} \aap, 669, A63, \dodoi{10.1051/0004-6361/202245004}

\bibitem[{V. {Bourrier} {et~al.}(2025){Bourrier}, {Steiner}, {Castro-Gonz{\'a}lez}, {Armstrong}, {Attia}, {Gill}, {Timmermans}, {Fernandez}, {Hawthorn}, {Triaud}, {Murgas}, {Palle}, {Chakraborty}, {Poppenhaeger}, {Lendl}, {Anderson}, {Bryant}, {Friden}, {Seidel}, {Zapatero Osorio}, {Eeles-Nolle}, {Lafarga}, {Lockley}, {Serrano Bell}, {Allart}, {Meech}, {Osborn}, {D{\'\i}az}, {Fetzner Keniger}, {Frame}, {Heitzmann}, {Ringham}, {Eggenberger}, {Alibert}, {Almenara}, {Leleu}, {Sousa}, {Mercier}, {Adibekyan}, {Battley}, {Delgado Mena}, {Dethier}, {Egger}, {Barkaoui}, {Bayliss}, {Burdanov}, {Ducrot}, {Ghachoui}, {Gillon}, {G{\'o}mez Maqueo Chew}, {Jehin}, {Pedersen}, {Pozuelos}, {Wheatley}, {Z{\'u}niga-Fern{\'a}ndez}, {Carteret}, {Cegla}, {Correia}, {Davis}, {Doyle}, {Ehrenreich}, {Hara}, {Lavie}, {Lillo-Box}, {Lovis}, {Petit}, {Santos}, {Scott}, {Venturini}, {Ahrer}, {Aigrain}, {Barros}, {Gillen}, {Luo}, {Mordasini}, {Al Moulla}, {Pepe}, \& {Pietrow}}]{Bourrier2025}
{Bourrier}, V., {Steiner}, M., {Castro-Gonz{\'a}lez}, A., {et~al.} 2025, \bibinfo{title}{{ATREIDES: I. Embarking on a trek across the exo-Neptunian landscape with the TOI-421 system},} \aap, 701, A190, \dodoi{10.1051/0004-6361/202554856}

\bibitem[{R. {Brahm} {et~al.}(2017){Brahm}, {Jord{\'a}n}, {Hartman}, \& {Bakos}}]{zaspe}
{Brahm}, R., {Jord{\'a}n}, A., {Hartman}, J., \& {Bakos}, G. 2017, \bibinfo{title}{{ZASPE: A Code to Measure Stellar Atmospheric Parameters and their Covariance from Spectra},} \mnras, 467, 971, \dodoi{10.1093/mnras/stx144}

\bibitem[{R. {Brahm} {et~al.}(2019){Brahm}, {Espinoza}, {Jord{\'a}n}, {Henning}, {Sarkis}, {Jones}, {D{\'\i}az}, {Jenkins}, {Vanzi}, {Zapata}, {Petrovich}, {Kossakowski}, {Rabus}, {Rojas}, \& {Torres}}]{Brahm2019}
{Brahm}, R., {Espinoza}, N., {Jord{\'a}n}, A., {et~al.} 2019, \bibinfo{title}{{HD 1397b: A Transiting Warm Giant Planet Orbiting A V = 7.8 mag Subgiant Star Discovered by TESS},} \aj, 158, 45, \dodoi{10.3847/1538-3881/ab279a}

\bibitem[{A. {Bressan} {et~al.}(2012){Bressan}, {Marigo}, {Girardi}, {Salasnich}, {Dal Cero}, {Rubele}, \& {Nanni}}]{parsec}
{Bressan}, A., {Marigo}, P., {Girardi}, L., {et~al.} 2012, \bibinfo{title}{{PARSEC: stellar tracks and isochrones with the PAdova and TRieste Stellar Evolution Code},} \mnras, 427, 127, \dodoi{10.1111/j.1365-2966.2012.21948.x}

\bibitem[{R.~P. {Butler} {et~al.}(1996){Butler}, {Marcy}, {Williams}, {McCarthy}, {Dosanjh}, \& {Vogt}}]{Butler1996}
{Butler}, R.~P., {Marcy}, G.~W., {Williams}, E., {et~al.} 1996, \bibinfo{title}{{Attaining Doppler Precision of 3 M s-1},} \pasp, 108, 500, \dodoi{10.1086/133755}

\bibitem[{D.~A. Caldwell {et~al.}(2020)Caldwell, Jenkins, \& Ting}]{TESS_30min}
Caldwell, D.~A., Jenkins, J.~M., \& Ting, E.~B. 2020, TESS Light Curves From Full Frame Images ("TESS-SPOC"), STScI/MAST, \dodoi{10.17909/T9-WPZ1-8S54}

\bibitem[{T.~L. {Campante} {et~al.}(2016){Campante}, {Lund}, {Kuszlewicz}, {Davies}, {Chaplin}, {Albrecht}, {Winn}, {Bedding}, {Benomar}, {Bossini}, {Handberg}, {Santos}, {Van Eylen}, {Basu}, {Christensen-Dalsgaard}, {Elsworth}, {Hekker}, {Hirano}, {Huber}, {Karoff}, {Kjeldsen}, {Lundkvist}, {North}, {Silva Aguirre}, {Stello}, \& {White}}]{Campante2016}
{Campante}, T.~L., {Lund}, M.~N., {Kuszlewicz}, J.~S., {et~al.} 2016, \bibinfo{title}{{Spin-Orbit Alignment of Exoplanet Systems: Ensemble Analysis Using Asteroseismology},} \apj, 819, 85, \dodoi{10.3847/0004-637X/819/1/85}

\bibitem[{J.~A. {Cardelli} {et~al.}(1989){Cardelli}, {Clayton}, \& {Mathis}}]{Cardelli1989}
{Cardelli}, J.~A., {Clayton}, G.~C., \& {Mathis}, J.~S. 1989, \bibinfo{title}{{The Relationship between Infrared, Optical, and Ultraviolet Extinction},} \apj, 345, 245, \dodoi{10.1086/167900}

\bibitem[{A. {Castro-Gonz{\'a}lez} {et~al.}(2026){Castro-Gonz{\'a}lez}, {Bourrier}, {Ehrenreich}, {Armstrong}, {Correia}, \& {Lendl}}]{Castro-Gonzalez2026}
{Castro-Gonz{\'a}lez}, A., {Bourrier}, V., {Ehrenreich}, D., {et~al.} 2026, \bibinfo{title}{{The Neptunian ridge as a natural outcome of high-eccentricity tidal migration},} \aap, 709, L17, \dodoi{10.1051/0004-6361/202659558}

\bibitem[{A. {Castro-Gonz{\'a}lez} {et~al.}(2024){Castro-Gonz{\'a}lez}, {Bourrier}, {Lillo-Box}, {Delisle}, {Armstrong}, {Barrado}, \& {Correia}}]{Castro-Gonzalez2024}
{Castro-Gonz{\'a}lez}, A., {Bourrier}, V., {Lillo-Box}, J., {et~al.} 2024, \bibinfo{title}{{Mapping the exo-Neptunian landscape: A ridge between the desert and savanna},} \aap, 689, A250, \dodoi{10.1051/0004-6361/202450957}

\bibitem[{A. {Cevallos Soto} \& Z. {Zhu}(2026){Cevallos Soto} \& {Zhu}}]{CevallosSoto2026}
{Cevallos Soto}, A., \& {Zhu}, Z. 2026, \bibinfo{title}{{Young Planets around Young Accreting Stars. I. Migration and Inner Stalling Orbits},} \apj, 998, 209, \dodoi{10.3847/1538-4357/ae355c}

\bibitem[{R.~I. {Dawson} \& J.~A. {Johnson}(2018){Dawson} \& {Johnson}}]{Dawson2018}
{Dawson}, R.~I., \& {Johnson}, J.~A. 2018, \bibinfo{title}{{Origins of Hot Jupiters},} \araa, 56, 175, \dodoi{10.1146/annurev-astro-081817-051853}

\bibitem[{J. {Dong} \& D. {Foreman-Mackey}(2023){Dong} \& {Foreman-Mackey}}]{Dong2023}
{Dong}, J., \& {Foreman-Mackey}, D. 2023, \bibinfo{title}{{A Hierarchical Bayesian Framework for Inferring the Stellar Obliquity Distribution},} \aj, 166, 112, \dodoi{10.3847/1538-3881/ace105}

\bibitem[{S. {Dong} {et~al.}(2018){Dong}, {Xie}, {Zhou}, {Zheng}, \& {Luo}}]{Dong2018}
{Dong}, S., {Xie}, J.-W., {Zhou}, J.-L., {Zheng}, Z., \& {Luo}, A. 2018, \bibinfo{title}{{LAMOST telescope reveals that Neptunian cousins of hot Jupiters are mostly single offspring of stars that are rich in heavy elements},} Proceedings of the National Academy of Science, 115, 266, \dodoi{10.1073/pnas.1711406115}

\bibitem[{E. {Dugan} {et~al.}(2025){Dugan}, {Wang}, {Heron}, {Bhaskar}, {Rice}, {Petrovich}, \& {Wang}}]{Dugan2025}
{Dugan}, E., {Wang}, X.-Y., {Heron}, A., {et~al.} 2025, \bibinfo{title}{{Early Evidence for Polar Orbits of Sub-Saturns around Hot Stars},} \apjl, 994, L23, \dodoi{10.3847/2041-8213/ae18c7}

\bibitem[{F. {Eeles-Nolle} \& D.~J. {Armstrong}(2025){Eeles-Nolle} \& {Armstrong}}]{Eeles-Nolle2025}
{Eeles-Nolle}, F., \& {Armstrong}, D.~J. 2025, \bibinfo{title}{{A high stellar multiplicity rate amongst TESS planet candidates in the Neptunian desert using Gaia DR3 astrometry},} \mnras, 541, 1419, \dodoi{10.1093/mnras/staf1072}

\bibitem[{K. {El-Badry} {et~al.}(2021){El-Badry}, {Rix}, \& {Heintz}}]{El-Badry2021}
{El-Badry}, K., {Rix}, H.-W., \& {Heintz}, T.~M. 2021, \bibinfo{title}{{A million binaries from Gaia eDR3: sample selection and validation of Gaia parallax uncertainties},} \mnras, 506, 2269, \dodoi{10.1093/mnras/stab323}

\bibitem[{N. {Espinoza} {et~al.}(2019){Espinoza}, {Kossakowski}, \& {Brahm}}]{juliet}
{Espinoza}, N., {Kossakowski}, D., \& {Brahm}, R. 2019, \bibinfo{title}{{juliet: a versatile modelling tool for transiting and non-transiting exoplanetary systems},} \mnras, 490, 2262, \dodoi{10.1093/mnras/stz2688}

\bibitem[{J.~I. {Espinoza-Retamal} {et~al.}(2023){Espinoza-Retamal}, {Brahm}, {Petrovich}, {Jord{\'a}n}, {Stef{\'a}nsson}, {Sedaghati}, {Hobson}, {Mu{\~n}oz}, {Boyle}, {Leiva}, \& {Suc}}]{Espinoza-Retamal2023b}
{Espinoza-Retamal}, J.~I., {Brahm}, R., {Petrovich}, C., {et~al.} 2023, \bibinfo{title}{{The Aligned Orbit of the Eccentric Proto Hot Jupiter TOI-3362b},} \apjl, 958, L20, \dodoi{10.3847/2041-8213/ad096d}

\bibitem[{J.~I. {Espinoza-Retamal} {et~al.}(2024){Espinoza-Retamal}, {Stef{\'a}nsson}, {Petrovich}, {Brahm}, {Jord{\'a}n}, {Sedaghati}, {Lucero}, {Pinto}, {Mu{\~n}oz}, {Boyle}, {Leiva}, \& {Suc}}]{Espinoza-Retamal2024}
{Espinoza-Retamal}, J.~I., {Stef{\'a}nsson}, G., {Petrovich}, C., {et~al.} 2024, \bibinfo{title}{{HATS-38 b and WASP-139 b Join a Growing Group of Hot Neptunes on Polar Orbits},} \aj, 168, 185, \dodoi{10.3847/1538-3881/ad70b8}

\bibitem[{J.~I. {Espinoza-Retamal} {et~al.}(2026{\natexlab{a}}){Espinoza-Retamal}, {Winn}, {Brahm}, {Petrovich}, {Stef{\'a}nsson}, {Bhaskar}, {Koo}, {Jord{\'a}n}, {Tala Pinto}, {Hobson}, {Veldhuis}, {Rojas}, {Teske}, {Butler}, {Crane}, {Shectman}, {Vissapragada}, {Boyle}, {Leiva}, \& {Suc}}]{Espinoza-Retamal2026}
{Espinoza-Retamal}, J.~I., {Winn}, J.~N., {Brahm}, R., {et~al.} 2026{\natexlab{a}}, \bibinfo{title}{{POSEIDON I: The Dynamical Origins of Transiting Neptunes},} \aj, 172, 57, \dodoi{10.3847/1538-3881/ae75da}

\bibitem[{J.~I. {Espinoza-Retamal} {et~al.}(2026{\natexlab{b}}){Espinoza-Retamal}, {Bhaskar}, {Winn}, {Petrovich}, {Brahm}, {Lammers}, {Stef{\'a}nsson}, {Koo}, {Jord{\'a}n}, \& {Rojas}}]{Espinoza-Retamal2026b}
{Espinoza-Retamal}, J.~I., {Bhaskar}, H., {Winn}, J.~N., {et~al.} 2026{\natexlab{b}}, \bibinfo{title}{{POSEIDON. II. The Antialigned Orbit of the Warm Neptune TOI-1710 A b},} \apjl, 1005, L15, \dodoi{10.3847/2041-8213/ae7711}

\bibitem[{D. {Foreman-Mackey} {et~al.}(2017){Foreman-Mackey}, {Agol}, {Angus}, \& {Ambikasaran}}]{celerite}
{Foreman-Mackey}, D., {Agol}, E., {Angus}, R., \& {Ambikasaran}, S. 2017, \bibinfo{title}{Fast and scalable Gaussian process modeling with applications to astronomical time series,} AJ, 154, 220, \dodoi{10.3847/1538-3881/aa9332}

\bibitem[{B.~J. {Fulton} {et~al.}(2018){Fulton}, {Petigura}, {Blunt}, \& {Sinukoff}}]{Fulton2018}
{Fulton}, B.~J., {Petigura}, E.~A., {Blunt}, S., \& {Sinukoff}, E. 2018, \bibinfo{title}{{RadVel: The Radial Velocity Modeling Toolkit},} \pasp, 130, 044504, \dodoi{10.1088/1538-3873/aaaaa8}

\bibitem[{ {Gaia Collaboration} {et~al.}(2021){Gaia Collaboration}, {Brown}, {Vallenari}, {Prusti}, {de Bruijne}, {Babusiaux}, {Biermann}, {Creevey}, {Evans}, {Eyer}, {Hutton}, {Jansen}, {Jordi}, {Klioner}, {Lammers}, {Lindegren}, {Luri}, {Mignard}, {Panem}, {Pourbaix}, {Randich}, {Sartoretti}, {Soubiran}, {Walton}, {Arenou}, {Bailer-Jones}, {Bastian}, {Cropper}, {Drimmel}, {Katz}, {Lattanzi}, {van Leeuwen}, {Bakker}, {Cacciari}, {Casta{\~n}eda}, {De Angeli}, {Ducourant}, {Fabricius}, {Fouesneau}, {Fr{\'e}mat}, {Guerra}, {Guerrier}, {Guiraud}, {Jean-Antoine Piccolo}, {Masana}, {Messineo}, {Mowlavi}, {Nicolas}, {Nienartowicz}, {Pailler}, {Panuzzo}, {Riclet}, {Roux}, {Seabroke}, {Sordo}, {Tanga}, {Th{\'e}venin}, {Gracia-Abril}, {Portell}, {Teyssier}, {Altmann}, {Andrae}, {Bellas-Velidis}, {Benson}, {Berthier}, {Blomme}, {Brugaletta}, {Burgess}, {Busso}, {Carry}, {Cellino}, {Cheek}, {Clementini}, {Damerdji}, {Davidson}, {Delchambre}, {Dell'Oro}, {Fern{\'a}ndez-Hern{\'a}ndez}, {Galluccio}, {Garc{\'\i}a-Lario},
  {Garcia-Reinaldos}, {Gonz{\'a}lez-N{\'u}{\~n}ez}, {Gosset}, {Haigron}, {Halbwachs}, {Hambly}, {Harrison}, {Hatzidimitriou}, {Heiter}, {Hern{\'a}ndez}, {Hestroffer}, {Hodgkin}, {Holl}, {Jan{\ss}en}, {Jevardat de Fombelle}, {Jordan}, {Krone-Martins}, {Lanzafame}, {L{\"o}ffler}, {Lorca}, {Manteiga}, {Marchal}, {Marrese}, {Moitinho}, {Mora}, {Muinonen}, {Osborne}, {Pancino}, {Pauwels}, {Petit}, {Recio-Blanco}, {Richards}, {Riello}, {Rimoldini}, {Robin}, {Roegiers}, {Rybizki}, {Sarro}, {Siopis}, {Smith}, {Sozzetti}, {Ulla}, {Utrilla}, {van Leeuwen}, {van Reeven}, {Abbas}, {Abreu Aramburu}, {Accart}, {Aerts}, {Aguado}, {Ajaj}, {Altavilla}, {{\'A}lvarez}, {{\'A}lvarez Cid-Fuentes}, {Alves}, {Anderson}, {Anglada Varela}, {Antoja}, {Audard}, {Baines}, {Baker}, {Balaguer-N{\'u}{\~n}ez}, {Balbinot}, {Balog}, {Barache}, {Barbato}, {Barros}, {Barstow}, {Bartolom{\'e}}, {Bassilana}, {Bauchet}, {Baudesson-Stella}, {Becciani}, {Bellazzini}, {Bernet}, {Bertone}, {Bianchi}, {Blanco-Cuaresma}, {Boch}, {Bombrun}, {Bossini},
  {Bouquillon}, {Bragaglia}, {Bramante}, {Breedt}, {Bressan}, {Brouillet}, {Bucciarelli}, {Burlacu}, {Busonero}, {Butkevich}, {Buzzi}, {Caffau}, {Cancelliere}, {C{\'a}novas}, {Cantat-Gaudin}, {Carballo}, {Carlucci}, {Carnerero}, {Carrasco}, {Casamiquela}, {Castellani}, {Castro-Ginard}, {Castro Sampol}, {Chaoul}, {Charlot}, {Chemin}, {Chiavassa}, {Cioni}, {Comoretto}, {Cooper}, {Cornez}, {Cowell}, {Crifo}, {Crosta}, {Crowley}, {Dafonte}, {Dapergolas}, {David}, \& {David}}]{GaiaEDR3}
{Gaia Collaboration}, {Brown}, A.~G.~A., {Vallenari}, A., {et~al.} 2021, \bibinfo{title}{{Gaia Early Data Release 3. Summary of the contents and survey properties},} \aap, 649, A1, \dodoi{10.1051/0004-6361/202039657}

\bibitem[{ {Gaia Collaboration} {et~al.}(2023){Gaia Collaboration}, {Vallenari}, {Brown}, {Prusti}, {de Bruijne}, {Arenou}, {Babusiaux}, {Biermann}, {Creevey}, {Ducourant}, {Evans}, {Eyer}, {Guerra}, {Hutton}, {Jordi}, {Klioner}, {Lammers}, {Lindegren}, {Luri}, {Mignard}, {Panem}, {Pourbaix}, {Randich}, {Sartoretti}, {Soubiran}, {Tanga}, {Walton}, {Bailer-Jones}, {Bastian}, {Drimmel}, {Jansen}, {Katz}, {Lattanzi}, {van Leeuwen}, {Bakker}, {Cacciari}, {Casta{\~n}eda}, {De Angeli}, {Fabricius}, {Fouesneau}, {Fr{\'e}mat}, {Galluccio}, {Guerrier}, {Heiter}, {Masana}, {Messineo}, {Mowlavi}, {Nicolas}, {Nienartowicz}, {Pailler}, {Panuzzo}, {Riclet}, {Roux}, {Seabroke}, {Sordo}, {Th{\'e}venin}, {Gracia-Abril}, {Portell}, {Teyssier}, {Altmann}, {Andrae}, {Audard}, {Bellas-Velidis}, {Benson}, {Berthier}, {Blomme}, {Burgess}, {Busonero}, {Busso}, {C{\'a}novas}, {Carry}, {Cellino}, {Cheek}, {Clementini}, {Damerdji}, {Davidson}, {de Teodoro}, {Nu{\~n}ez Campos}, {Delchambre}, {Dell'Oro}, {Esquej},
  {Fern{\'a}ndez-Hern{\'a}ndez}, {Fraile}, {Garabato}, {Garc{\'\i}a-Lario}, {Gosset}, {Haigron}, {Halbwachs}, {Hambly}, {Harrison}, {Hern{\'a}ndez}, {Hestroffer}, {Hodgkin}, {Holl}, {Jan{\ss}en}, {Jevardat de Fombelle}, {Jordan}, {Krone-Martins}, {Lanzafame}, {L{\"o}ffler}, {Marchal}, {Marrese}, {Moitinho}, {Muinonen}, {Osborne}, {Pancino}, {Pauwels}, {Recio-Blanco}, {Reyl{\'e}}, {Riello}, {Rimoldini}, {Roegiers}, {Rybizki}, {Sarro}, {Siopis}, {Smith}, {Sozzetti}, {Utrilla}, {van Leeuwen}, {Abbas}, {{\'A}brah{\'a}m}, {Abreu Aramburu}, {Aerts}, {Aguado}, {Ajaj}, {Aldea-Montero}, {Altavilla}, {{\'A}lvarez}, {Alves}, {Anders}, {Anderson}, {Anglada Varela}, {Antoja}, {Baines}, {Baker}, {Balaguer-N{\'u}{\~n}ez}, {Balbinot}, {Balog}, {Barache}, {Barbato}, {Barros}, {Barstow}, {Bartolom{\'e}}, {Bassilana}, {Bauchet}, {Becciani}, {Bellazzini}, {Berihuete}, {Bernet}, {Bertone}, {Bianchi}, {Binnenfeld}, {Blanco-Cuaresma}, {Blazere}, {Boch}, {Bombrun}, {Bossini}, {Bouquillon}, {Bragaglia}, {Bramante}, {Breedt},
  {Bressan}, {Brouillet}, {Brugaletta}, {Bucciarelli}, {Burlacu}, {Butkevich}, {Buzzi}, {Caffau}, {Cancelliere}, {Cantat-Gaudin}, {Carballo}, {Carlucci}, {Carnerero}, {Carrasco}, {Casamiquela}, {Castellani}, {Castro-Ginard}, {Chaoul}, {Charlot}, {Chemin}, {Chiaramida}, {Chiavassa}, {Chornay}, {Comoretto}, {Contursi}, {Cooper}, {Cornez}, {Cowell}, {Crifo}, {Cropper}, {Crosta}, {Crowley}, {Dafonte}, {Dapergolas}, {David}, {David}, {de Laverny}, {De Luise}, \& {De March}}]{GaiaDR3}
{Gaia Collaboration}, {Vallenari}, A., {Brown}, A.~G.~A., {et~al.} 2023, \bibinfo{title}{{Gaia Data Release 3. Summary of the content and survey properties},} \aap, 674, A1, \dodoi{10.1051/0004-6361/202243940}

\bibitem[{B.~S. {Gaudi} \& J.~N. {Winn}(2007){Gaudi} \& {Winn}}]{Gaudi2007}
{Gaudi}, B.~S., \& {Winn}, J.~N. 2007, \bibinfo{title}{{Prospects for the Characterization and Confirmation of Transiting Exoplanets via the Rossiter-McLaughlin Effect},} \apj, 655, 550, \dodoi{10.1086/509910}

\bibitem[{S. {Giacalone} \& C.~D. {Dressing}(2025){Giacalone} \& {Dressing}}]{Giacalone2025}
{Giacalone}, S., \& {Dressing}, C.~D. 2025, \bibinfo{title}{{Small and Close-in Planets are Uncommon Around A-type Stars},} \aj, 169, 45, \dodoi{10.3847/1538-3881/ad9587}

\bibitem[{S. {Giacalone} {et~al.}(2022){Giacalone}, {Dressing}, {Garc{\'\i}a Mu{\~n}oz}, {Hooton}, {Stassun}, {Quinn}, {Zhou}, {Ziegler}, {Vanderspek}, {Latham}, {Seager}, {Winn}, {Jenkins}, {Brice{\~n}o}, {Huang}, {Rodriguez}, {Shporer}, {Mann}, {Watanabe}, \& {Wohler}}]{Giacalone2022}
{Giacalone}, S., {Dressing}, C.~D., {Garc{\'\i}a Mu{\~n}oz}, A., {et~al.} 2022, \bibinfo{title}{{HD 56414 b: A Warm Neptune Transiting an A-type Star},} \apjl, 935, L10, \dodoi{10.3847/2041-8213/ac80f4}

\bibitem[{S.~R. {Gibson} {et~al.}(2018){Gibson}, {Howard}, {Roy}, {Smith}, {Halverson}, {Edelstein}, {Kassis}, {Wishnow}, {Raffanti}, {Allen}, {Chin}, {Coutts}, {Cowley}, {Curtis}, {Deich}, {Feger}, {Finstad}, {Gurevich}, {Ishikawa}, {James}, {Jhoti}, {Lanclos}, {Lilley}, {Miller}, {Milner}, {Payne}, {Rider}, {Rockosi}, {Sandford}, {Schwab}, {Seifahrt}, {Sirk}, {Smith}, {Stuermer}, {Weisfeiler}, {Wilcox}, {Vandenberg}, \& {Wizinowich}}]{Gibson2018}
{Gibson}, S.~R., {Howard}, A.~W., {Roy}, A., {et~al.} 2018, \bibinfo{title}{{Keck Planet Finder: preliminary design},} in Society of Photo-Optical Instrumentation Engineers (SPIE) Conference Series, Vol. 10702, Ground-based and Airborne Instrumentation for Astronomy VII, ed. C.~J. {Evans}, L.~{Simard}, \& H.~{Takami}, 107025X, \dodoi{10.1117/12.2311565}

\bibitem[{S.~R. {Gibson} {et~al.}(2020){Gibson}, {Howard}, {Rider}, {Roy}, {Edelstein}, {Kassis}, {Grillo}, {Halverson}, {Sirk}, {Smith}, {Allen}, {Baker}, {Beichman}, {Berriman}, {Brown}, {Casey}, {Chin}, {Coutts}, {Cowley}, {Deich}, {Feger}, {Fulton}, {Gers}, {Gurevich}, {Ishikawa}, {James}, {Jelinsky}, {Kaye}, {Lanclos}, {Li}, {Lilley}, {McCarney}, {Miller}, {Milner}, {O'Hanlon}, {Pember}, {Raffanti}, {Rockosi}, {Rubenzahl}, {Rumph}, {Sandford}, {Savage}, {Schwab}, {Seifahrt}, {Shaum}, {Smith}, {Stuermer}, {Thorne}, {Vandenberg}, {Von Boeckmann}, {Wang}, {Wang}, {Weisfeiler}, {Wilcox}, {Wishnow}, {Wizinowich}, {Wold}, \& {Wolfenberger}}]{Gibson2020}
{Gibson}, S.~R., {Howard}, A.~W., {Rider}, K., {et~al.} 2020, \bibinfo{title}{{Keck Planet Finder: design updates},} in Society of Photo-Optical Instrumentation Engineers (SPIE) Conference Series, Vol. 11447, Ground-based and Airborne Instrumentation for Astronomy VIII, ed. C.~J. {Evans}, J.~J. {Bryant}, \& K.~{Motohara}, 1144742, \dodoi{10.1117/12.2561783}

\bibitem[{S.~R. {Gibson} {et~al.}(2024){Gibson}, {Howard}, {Rider}, {Halverson}, {Roy}, {Baker}, {Edelstein}, {Smith}, {Fulton}, {Walawender}, {Brodheim}, {Brown}, {Chan}, {Dai}, {Deich}, {Gottschalk}, {Grillo}, {Hale}, {Hill}, {Holden}, {Householder}, {Isaacson}, {Ishikawa}, {Jelinsky}, {Kassis}, {Kaye}, {Laher}, {Lanclos}, {Lee}, {Lilley}, {McCarney}, {Miller}, {Payne}, {Petigura}, {Poppett}, {Raffanti}, {Rubenzahl}, {Sandford}, {Schwab}, {Shaum}, {Sirk}, {Smith}, {Thorne}, {Valliant}, {Vandenberg}, {Wang}, {Wishnow}, {Wold}, {Yeh}, {Baca}, {Beichman}, {Berriman}, {Brown}, {Casey}, {Chin}, {Chong}, {Cowley}, {Devenot}, {Elwir}, {Finstad}, {Fraysse}, {James}, {Jhoti}, {Killian}, {Levine}, {Li}, {Marin}, {Milner}, {Nance}, {O'Hanlon}, {Orr}, {Ortiz-Soto}, {Payne}, {Pember}, {Raskin}, {Savage}, {Seifahrt}, {Smith}, {Storesund}, {St{\"u}rmer}, {Suominen}, {Tehero}, {Von Boeckmann}, {Wages}, {Weisfeiler}, {Wilcox}, {Wizinowich}, \& {Wolfenberger}}]{Gibson2024}
{Gibson}, S.~R., {Howard}, A.~W., {Rider}, K., {et~al.} 2024, \bibinfo{title}{{System design of the Keck Planet Finder},} in Society of Photo-Optical Instrumentation Engineers (SPIE) Conference Series, Vol. 13096, Ground-based and Airborne Instrumentation for Astronomy X, ed. J.~J. {Bryant}, K.~{Motohara}, \& J.~R.~D. {Vernet}, 1309609, \dodoi{10.1117/12.3017841}

\bibitem[{P. {Goldreich} \& S. {Soter}(1966){Goldreich} \& {Soter}}]{Goldreich66}
{Goldreich}, P., \& {Soter}, S. 1966, \bibinfo{title}{{Q in the Solar System},} \icarus, 5, 375, \dodoi{10.1016/0019-1035(66)90051-0}

\bibitem[{P. {Goldreich} \& S. {Tremaine}(1980){Goldreich} \& {Tremaine}}]{Goldreich1980}
{Goldreich}, P., \& {Tremaine}, S. 1980, \bibinfo{title}{{Disk-satellite interactions.},} \apj, 241, 425, \dodoi{10.1086/158356}

\bibitem[{D.~F. {Gray}(1984){Gray}}]{Gray1984}
{Gray}, D.~F. 1984, \bibinfo{title}{{Measurements of rotation and turbulence in F, G and K dwarfs.},} \apj, 281, 719, \dodoi{10.1086/162149}

\bibitem[{S. {Grouffal} {et~al.}(2022){Grouffal}, {Santerne}, {Bourrier}, {Dumusque}, {Triaud}, {Malavolta}, {Kunovac}, {Armstrong}, {Attia}, {Barros}, {Boisse}, {Deleuil}, {Demangeon}, {Dressing}, {Figueira}, {Lillo-Box}, {Mortier}, {Nardiello}, {Santos}, \& {Sousa}}]{Grouffal2022}
{Grouffal}, S., {Santerne}, A., {Bourrier}, V., {et~al.} 2022, \bibinfo{title}{{Rossiter-McLaughlin detection of the 9-month period transiting exoplanet HIP41378 d},} \aap, 668, A172, \dodoi{10.1051/0004-6361/202244182}

\bibitem[{S. {Grouffal} {et~al.}(2025){Grouffal}, {Santerne}, {Bourrier}, {Kunovac}, {Dressing}, {Akinsanmi}, {Armstrong}, {Baliwal}, {Balsalobre-Ruza}, {Barros}, {Bayliss}, {Crossfield}, {Demangeon}, {Dumusque}, {Giacalone}, {Harada}, {Isaacson}, {Kellermann}, {Lillo-Box}, {Llama}, {Mortier}, {Palle}, {Rajpurohit}, {Rice}, {Santos}, {Seidel}, {Sharma}, {Sousa}, {Thomas}, {Turtelboom}, {Udry}, \& {Wheatley}}]{Grouffal2025}
{Grouffal}, S., {Santerne}, A., {Bourrier}, V., {et~al.} 2025, \bibinfo{title}{{The star HIP 41378 potentially misaligned with its cohort of long-period planets},} \aap, 701, A173, \dodoi{10.1051/0004-6361/202555487}

\bibitem[{T. {Hallatt} \& S. {Millholland}(2026{\natexlab{a}}){Hallatt} \& {Millholland}}]{Hallatt2026b}
{Hallatt}, T., \& {Millholland}, S. 2026{\natexlab{a}}, \bibinfo{title}{{Coupled Planetary Interior and Tidal Evolution},} \apj, 997, 138, \dodoi{10.3847/1538-4357/ae129d}

\bibitem[{T. {Hallatt} \& S. {Millholland}(2026{\natexlab{b}}){Hallatt} \& {Millholland}}]{Hallatt2025}
{Hallatt}, T., \& {Millholland}, S. 2026{\natexlab{b}}, \bibinfo{title}{{Shedding Light on Desert Dwellers},} \apj, 997, 139, \dodoi{10.3847/1538-4357/adfb75}

\bibitem[{T. {Hallatt} {et~al.}(2026){Hallatt}, {Owen}, \& {Millholland}}]{Hallatt2026}
{Hallatt}, T., {Owen}, J.~E., \& {Millholland}, S. 2026, \bibinfo{title}{{Revealing the Origin of Desert Dwellers via Stellar Obliquities},} \apjl, 1005, L17, \dodoi{10.3847/2041-8213/ae6fb5}

\bibitem[{S. {Halverson} {et~al.}(2016){Halverson}, {Terrien}, {Mahadevan}, {Roy}, {Bender}, {Stef{\'a}nsson}, {Monson}, {Levi}, {Hearty}, {Blake}, {McElwain}, {Schwab}, {Ramsey}, {Wright}, {Wang}, {Gong}, \& {Roberston}}]{Halverson2016}
{Halverson}, S., {Terrien}, R., {Mahadevan}, S., {et~al.} 2016, \bibinfo{title}{{A comprehensive radial velocity error budget for next generation Doppler spectrometers},} in Society of Photo-Optical Instrumentation Engineers (SPIE) Conference Series, Vol. 9908, Ground-based and Airborne Instrumentation for Astronomy VI, ed. C.~J. {Evans}, L.~{Simard}, \& H.~{Takami}, 99086P, \dodoi{10.1117/12.2232761}

\bibitem[{L.~B. {Handley} {et~al.}(2024{\natexlab{a}}){Handley}, {Petigura}, \& {Mi{\v{s}}i{\'c}}}]{Handley2024a}
{Handley}, L.~B., {Petigura}, E.~A., \& {Mi{\v{s}}i{\'c}}, V.~V. 2024{\natexlab{a}}, \bibinfo{title}{{Solving the Traveling Telescope Problem with Mixed-integer Linear Programming},} \aj, 167, 33, \dodoi{10.3847/1538-3881/ad0dfb}

\bibitem[{L.~B. {Handley} {et~al.}(2024{\natexlab{b}}){Handley}, {Petigura}, {Mi{\v{s}}i{\'c}}, {Lubin}, \& {Isaacson}}]{Handley2024b}
{Handley}, L.~B., {Petigura}, E.~A., {Mi{\v{s}}i{\'c}}, V.~V., {Lubin}, J., \& {Isaacson}, H. 2024{\natexlab{b}}, \bibinfo{title}{{Automated Scheduling of Doppler Exoplanet Observations at Keck Observatory},} \aj, 167, 122, \dodoi{10.3847/1538-3881/ad1ff7}

\bibitem[{L.~B. {Handley} {et~al.}(2026){Handley}, {Howard}, {Dai}, {Rubenzahl}, {Giacalone}, {Isaacson}, {Ong}, {Carmichael}, {Li}, {Lubin}, {Premnath}, {Rogers}, {Nagarajan}, {Gilbert}, {Fulton}, {Gibson}, {Roy}, {Edelstein}, \& {Smith}}]{Handley2026}
{Handley}, L.~B., {Howard}, A.~W., {Dai}, F., {et~al.} 2026, \bibinfo{title}{{The KPF-SLOPE Survey - Small, Compact Multi-Planet Systems Appear Spin-Orbit Aligned},} arXiv e-prints, arXiv:2603.23713, \dodoi{10.48550/arXiv.2603.23713}

\bibitem[{K.~K. {Hardegree-Ullman} {et~al.}(2025){Hardegree-Ullman}, {Bergsten}, {Christiansen}, {Zink}, {Bhure}, {Boley}, {Fernandes}, {Giacalone}, \& {Karpoor}}]{Hardegree-Ullman2025}
{Hardegree-Ullman}, K.~K., {Bergsten}, G.~J., {Christiansen}, J.~L., {et~al.} 2025, \bibinfo{title}{{Scaling K2. VIII. Short-period Sub-Neptune Occurrence Rates Peak Around Early-type M Dwarfs},} \aj, 170, 183, \dodoi{10.3847/1538-3881/adf633}

\bibitem[{J.~D. {Hartman} {et~al.}(2015){Hartman}, {Bakos}, {Buchhave}, {Torres}, {Latham}, {Kov{\'a}cs}, {Bhatti}, {Csubry}, {de Val-Borro}, {Penev}, {Huang}, {B{\'e}ky}, {Bieryla}, {Quinn}, {Howard}, {Marcy}, {Johnson}, {Isaacson}, {Fischer}, {Noyes}, {Falco}, {Esquerdo}, {Knox}, {Hinz}, {L{\'a}z{\'a}r}, {Papp}, \& {S{\'a}ri}}]{Hartman2015}
{Hartman}, J.~D., {Bakos}, G.~{\'A}., {Buchhave}, L.~A., {et~al.} 2015, \bibinfo{title}{{HAT-P-57b: A Short-period Giant Planet Transiting a Bright Rapidly Rotating A8V Star Confirmed Via Doppler Tomography},} \aj, 150, 197, \dodoi{10.1088/0004-6256/150/6/197}

\bibitem[{Y. {He} {et~al.}(2026){He}, {Bitsch}, {Houge}, {Williams}, \& {Ogihara}}]{He2026}
{He}, Y., {Bitsch}, B., {Houge}, A., {Williams}, J., \& {Ogihara}, M. 2026, \bibinfo{title}{{The majority of hot Jupiters formed beyond the water ice line},} arXiv e-prints, arXiv:2607.15144, \dodoi{10.48550/arXiv.2607.15144}

\bibitem[{G. {H{\'e}brard} {et~al.}(2008){H{\'e}brard}, {Bouchy}, {Pont}, {Loeillet}, {Rabus}, {Bonfils}, {Moutou}, {Boisse}, {Delfosse}, {Desort}, {Eggenberger}, {Ehrenreich}, {Forveille}, {Lagrange}, {Lovis}, {Mayor}, {Pepe}, {Perrier}, {Queloz}, {Santos}, {S{\'e}gransan}, {Udry}, \& {Vidal-Madjar}}]{Hebrard2008}
{H{\'e}brard}, G., {Bouchy}, F., {Pont}, F., {et~al.} 2008, \bibinfo{title}{{Misaligned spin-orbit in the XO-3 planetary system?},} \aap, 488, 763, \dodoi{10.1051/0004-6361:200810056}

\bibitem[{T. {Hirano} {et~al.}(2010){Hirano}, {Suto}, {Taruya}, {Narita}, {Sato}, {Johnson}, \& {Winn}}]{Hirano2010}
{Hirano}, T., {Suto}, Y., {Taruya}, A., {et~al.} 2010, \bibinfo{title}{{Analytic Description of the Rossiter-Mclaughlin Effect for Transiting Exoplanets: Cross-Correlation Method and Comparison with Simulated Data},} \apj, 709, 458, \dodoi{10.1088/0004-637X/709/1/458}

\bibitem[{M. {Hjorth} {et~al.}(2021){Hjorth}, {Albrecht}, {Hirano}, {Winn}, {Dawson}, {Zanazzi}, {Knudstrup}, \& {Sato}}]{Hjorth2021}
{Hjorth}, M., {Albrecht}, S., {Hirano}, T., {et~al.} 2021, \bibinfo{title}{{A backward-spinning star with two coplanar planets},} Proceedings of the National Academy of Science, 118, e2017418118, \dodoi{10.1073/pnas.2017418118}

\bibitem[{A.~W. {Howard} {et~al.}(2010{\natexlab{a}}){Howard}, {Marcy}, {Johnson}, {Fischer}, {Wright}, {Isaacson}, {Valenti}, {Anderson}, {Lin}, \& {Ida}}]{Howard2010science}
{Howard}, A.~W., {Marcy}, G.~W., {Johnson}, J.~A., {et~al.} 2010{\natexlab{a}}, \bibinfo{title}{{The Occurrence and Mass Distribution of Close-in Super-Earths, Neptunes, and Jupiters},} Science, 330, 653, \dodoi{10.1126/science.1194854}

\bibitem[{A.~W. {Howard} {et~al.}(2010{\natexlab{b}}){Howard}, {Johnson}, {Marcy}, {Fischer}, {Wright}, {Bernat}, {Henry}, {Peek}, {Isaacson}, {Apps}, {Endl}, {Cochran}, {Valenti}, {Anderson}, \& {Piskunov}}]{Howard2010}
{Howard}, A.~W., {Johnson}, J.~A., {Marcy}, G.~W., {et~al.} 2010{\natexlab{b}}, \bibinfo{title}{{The California Planet Survey. I. Four New Giant Exoplanets},} \apj, 721, 1467, \dodoi{10.1088/0004-637X/721/2/1467}

\bibitem[{J.~D. {Hunter}(2007){Hunter}}]{matplotlib}
{Hunter}, J.~D. 2007, \bibinfo{title}{{Matplotlib: A 2D Graphics Environment},} Computing in Science and Engineering, 9, 90, \dodoi{10.1109/MCSE.2007.55}

\bibitem[{P. {Hut}(1981){Hut}}]{Hut81}
{Hut}, P. 1981, \bibinfo{title}{{Tidal evolution in close binary systems.},} \aap, 99, 126

\bibitem[{J.~M. {Jenkins} {et~al.}(2016){Jenkins}, {Twicken}, {McCauliff}, {Campbell}, {Sanderfer}, {Lung}, {Mansouri-Samani}, {Girouard}, {Tenenbaum}, {Klaus}, {Smith}, {Caldwell}, {Chacon}, {Henze}, {Heiges}, {Latham}, {Morgan}, {Swade}, {Rinehart}, \& {Vanderspek}}]{spoc}
{Jenkins}, J.~M., {Twicken}, J.~D., {McCauliff}, S., {et~al.} 2016, \bibinfo{title}{{The TESS science processing operations center},} in Society of Photo-Optical Instrumentation Engineers (SPIE) Conference Series, Vol. 9913, Software and Cyberinfrastructure for Astronomy IV, ed. G.~{Chiozzi} \& J.~C. {Guzman}, 99133E, \dodoi{10.1117/12.2233418}

\bibitem[{S. {Kanodia} {et~al.}(2018){Kanodia}, {Mahadevan}, {Ramsey}, {Stefansson}, {Monson}, {Hearty}, {Blakeslee}, {Lubar}, {Bender}, {Ninan}, {Sterner}, {Roy}, {Halverson}, \& {Robertson}}]{Kanodia2018}
{Kanodia}, S., {Mahadevan}, S., {Ramsey}, L.~W., {et~al.} 2018, \bibinfo{title}{{Overview of the spectrometer optical fiber feed for the habitable-zone planet finder},} in Society of Photo-Optical Instrumentation Engineers (SPIE) Conference Series, Vol. 10702, Ground-based and Airborne Instrumentation for Astronomy VII, ed. C.~J. {Evans}, L.~{Simard}, \& H.~{Takami}, 107026Q, \dodoi{10.1117/12.2313491}

\bibitem[{D.~M. {Kipping}(2013{\natexlab{a}}){Kipping}}]{Kipping13}
{Kipping}, D.~M. 2013{\natexlab{a}}, \bibinfo{title}{{Efficient, uninformative sampling of limb darkening coefficients for two-parameter laws},} \mnras, 435, 2152, \dodoi{10.1093/mnras/stt1435}

\bibitem[{D.~M. {Kipping}(2013{\natexlab{b}}){Kipping}}]{Kipping2013_ecc}
{Kipping}, D.~M. 2013{\natexlab{b}}, \bibinfo{title}{{Parametrizing the exoplanet eccentricity distribution with the beta distribution.},} \mnras, 434, L51, \dodoi{10.1093/mnrasl/slt075}

\bibitem[{E. {Knudstrup} {et~al.}(2024){Knudstrup}, {Albrecht}, {Winn}, {Gandolfi}, {Zanazzi}, {Persson}, {Fridlund}, {Marcussen}, {Chontos}, {Keniger}, {Eisner}, {Bieryla}, {Isaacson}, {Howard}, {Hirsch}, {Murgas}, {Narita}, {Palle}, {Kawai}, \& {Baker}}]{Knudstrup2024}
{Knudstrup}, E., {Albrecht}, S.~H., {Winn}, J.~N., {et~al.} 2024, \bibinfo{title}{{Obliquities of exoplanet host stars: Nineteen new and updated measurements, and trends in the sample of 205 measurements},} \aap, 690, A379, \dodoi{10.1051/0004-6361/202450627}

\bibitem[{R.~P. {Kraft}(1967){Kraft}}]{Kraft1967}
{Kraft}, R.~P. 1967, \bibinfo{title}{{Studies of Stellar Rotation. V. The Dependence of Rotation on Age among Solar-Type Stars},} \apj, 150, 551, \dodoi{10.1086/149359}

\bibitem[{L. {Kreidberg}(2015){Kreidberg}}]{batman}
{Kreidberg}, L. 2015, \bibinfo{title}{{batman: BAsic Transit Model cAlculatioN in Python},} \pasp, 127, 1161, \dodoi{10.1086/683602}

\bibitem[{N.~T. {Kurtovic} {et~al.}(2026){Kurtovic}, {Flores-Rivera}, {Perez}, {Vioque}, {Benisty}, {Alarc{\'o}n}, {Barraza-Alfaro}, {Curone}, {Doi}, {Grant}, {Jiang}, {Kataoka}, {Long}, {Ribas}, {Sierra}, {Stapper}, {Temmink}, \& {Zagar{\'\i}a}}]{Kurtovic2026}
{Kurtovic}, N.~T., {Flores-Rivera}, L., {Perez}, L.~M., {et~al.} 2026, \bibinfo{title}{{An archival summary: 15 years of ALMA observations on disks and planet formation},} arXiv e-prints, arXiv:2605.30023, \dodoi{10.48550/arXiv.2605.30023}

\bibitem[{M. {Lafarga} {et~al.}(2026){Lafarga}, {Espinoza-Retamal}, {Cegla}, {Stef{\~A}{\textexclamdown}nsson}, {Freckelton}, {Mortier}, {Gill}, {Ahrer}, {Anderson}, {Armstrong}, {Bean}, {Bourrier}, {Brady}, {Brogi}, {Bryant}, {Burleigh}, {Doyle}, {Jenkins}, {Kasper}, {Luo}, {Mancini}, {Moyano}, {Saha}, {Southworth}, {Veras}, {Vines}, {Wheatley}, \& {Winn}}]{Lafarga2026}
{Lafarga}, M., {Espinoza-Retamal}, J.~I., {Cegla}, H.~M., {et~al.} 2026, \bibinfo{title}{{The Neptunian ridge planet WASP-156 b does not have a polar orbit},} \mnras, \dodoi{10.1093/mnras/stag996}

\bibitem[{ {Lightkurve Collaboration} {et~al.}(2018){Lightkurve Collaboration}, {Cardoso}, {Hedges}, {Gully-Santiago}, {Saunders}, {Cody}, {Barclay}, {Hall}, {Sagear}, {Turtelboom}, {Zhang}, {Tzanidakis}, {Mighell}, {Coughlin}, {Bell}, {Berta-Thompson}, {Williams}, {Dotson}, \& {Barentsen}}]{lightkurve}
{Lightkurve Collaboration}, {Cardoso}, J.~V.~d.~M., {Hedges}, C., {et~al.} 2018, {Lightkurve: Kepler and TESS time series analysis in Python},, Astrophysics Source Code Library \doeprint{1812.013}

\bibitem[{S.~J. {Lilley} {et~al.}(2022){Lilley}, {Rider}, {Thorne}, {Kassis}, {Gibson}, {Howard}, {Lanclos}, \& {Walawender}}]{Lilley2022}
{Lilley}, S.~J., {Rider}, K., {Thorne}, J., {et~al.} 2022, \bibinfo{title}{{A fiber injection unit for the Keck Planet Finder: opto-mechanical design},} in Society of Photo-Optical Instrumentation Engineers (SPIE) Conference Series, Vol. 12184, Ground-based and Airborne Instrumentation for Astronomy IX, ed. C.~J. {Evans}, J.~J. {Bryant}, \& K.~{Motohara}, 121844K, \dodoi{10.1117/12.2628818}

\bibitem[{D.~N.~C. {Lin} \& J. {Papaloizou}(1986){Lin} \& {Papaloizou}}]{Lin1986}
{Lin}, D.~N.~C., \& {Papaloizou}, J. 1986, \bibinfo{title}{{On the Tidal Interaction between Protoplanets and the Protoplanetary Disk. III. Orbital Migration of Protoplanets},} \apj, 309, 846, \dodoi{10.1086/164653}

\bibitem[{J. {Lubin} {et~al.}(2026){Lubin}, {Petigura}, {Mi{\v{s}}i{\'c}}, {Van Zandt}, \& {Handley}}]{Lubin2026}
{Lubin}, J., {Petigura}, E.~A., {Mi{\v{s}}i{\'c}}, V.~V., {Van Zandt}, J., \& {Handley}, L.~B. 2026, \bibinfo{title}{{AstroQ: Automated Scheduling of Cadenced Astronomical Observations},} \aj, 171, 85, \dodoi{10.3847/1538-3881/ae2a2e}

\bibitem[{G. {Mandushev} {et~al.}(2005){Mandushev}, {Torres}, {Latham}, {Charbonneau}, {Alonso}, {White}, {Stefanik}, {Dunham}, {Brown}, \& {O'Donovan}}]{Mandushev2005}
{Mandushev}, G., {Torres}, G., {Latham}, D.~W., {et~al.} 2005, \bibinfo{title}{{The Challenge of Wide-Field Transit Surveys: The Case of GSC 01944-02289},} \apj, 621, 1061, \dodoi{10.1086/427727}

\bibitem[{D. McLaughlin(1924)McLaughlin}]{McLaughlin1924}
McLaughlin, D. 1924, \bibinfo{title}{Some results of a spectrographic study of the Algol system.,} Astrophysical Journal, 60, 22-31 (1924), 60

\bibitem[{G.~D. {Mulders} {et~al.}(2015){Mulders}, {Pascucci}, \& {Apai}}]{Mulders2015}
{Mulders}, G.~D., {Pascucci}, I., \& {Apai}, D. 2015, \bibinfo{title}{{A Stellar-mass-dependent Drop in Planet Occurrence Rates},} \apj, 798, 112, \dodoi{10.1088/0004-637X/798/2/112}

\bibitem[{U. {Munari} {et~al.}(2014){Munari}, {Henden}, {Frigo}, {Zwitter}, {Bienaym{\'e}}, {Bland-Hawthorn}, {Boeche}, {Freeman}, {Gibson}, {Gilmore}, {Grebel}, {Helmi}, {Kordopatis}, {Levine}, {Navarro}, {Parker}, {Reid}, {Seabroke}, {Siebert}, {Siviero}, {Smith}, {Steinmetz}, {Templeton}, {Terrell}, {Welch}, {Williams}, \& {Wyse}}]{apass}
{Munari}, U., {Henden}, A., {Frigo}, A., {et~al.} 2014, \bibinfo{title}{{APASS Landolt-Sloan BVgri Photometry of RAVE Stars. I. Data, Effective Temperatures, and Reddenings},} \aj, 148, 81, \dodoi{10.1088/0004-6256/148/5/81}

\bibitem[{G.~I. {Ogilvie} \& D.~N.~C. {Lin}(2007){Ogilvie} \& {Lin}}]{Ogilvie07}
{Ogilvie}, G.~I., \& {Lin}, D.~N.~C. 2007, \bibinfo{title}{{Tidal Dissipation in Rotating Solar-Type Stars},} \apj, 661, 1180, \dodoi{10.1086/515435}

\bibitem[{S. {Perruchot} {et~al.}(2008){Perruchot}, {Kohler}, {Bouchy}, {Richaud}, {Richaud}, {Moreaux}, {Merzougui}, {Sottile}, {Hill}, {Knispel}, {Regal}, {Meunier}, {Ilovaisky}, {Le Coroller}, {Gillet}, {Schmitt}, {Pepe}, {Fleury}, {Sosnowska}, {Vors}, {M{\'e}gevand}, {Blanc}, {Carol}, {Point}, {Laloge}, \& {Brunel}}]{Perruchot2008}
{Perruchot}, S., {Kohler}, D., {Bouchy}, F., {et~al.} 2008, \bibinfo{title}{{The SOPHIE spectrograph: design and technical key-points for high throughput and high stability},} in Society of Photo-Optical Instrumentation Engineers (SPIE) Conference Series, Vol. 7014, Ground-based and Airborne Instrumentation for Astronomy II, ed. I.~S. {McLean} \& M.~M. {Casali}, 70140J, \dodoi{10.1117/12.787379}

\bibitem[{C. {Petrovich}(2015){Petrovich}}]{Petrovich2015}
{Petrovich}, C. 2015, \bibinfo{title}{{Hot Jupiters from Coplanar High-eccentricity Migration},} \apj, 805, 75, \dodoi{10.1088/0004-637X/805/1/75}

\bibitem[{D. {Queloz} {et~al.}(2000){Queloz}, {Eggenberger}, {Mayor}, {Perrier}, {Beuzit}, {Naef}, {Sivan}, \& {Udry}}]{Queloz2000}
{Queloz}, D., {Eggenberger}, A., {Mayor}, M., {et~al.} 2000, \bibinfo{title}{{Detection of a spectroscopic transit by the planet orbiting the star HD209458},} \aap, 359, L13, \dodoi{10.48550/arXiv.astro-ph/0006213}

\bibitem[{G.~R. {Ricker} {et~al.}(2015){Ricker}, {Winn}, {Vanderspek}, {Latham}, {Bakos}, {Bean}, {Berta-Thompson}, {Brown}, {Buchhave}, {Butler}, {Butler}, {Chaplin}, {Charbonneau}, {Christensen-Dalsgaard}, {Clampin}, {Deming}, {Doty}, {De Lee}, {Dressing}, {Dunham}, {Endl}, {Fressin}, {Ge}, {Henning}, {Holman}, {Howard}, {Ida}, {Jenkins}, {Jernigan}, {Johnson}, {Kaltenegger}, {Kawai}, {Kjeldsen}, {Laughlin}, {Levine}, {Lin}, {Lissauer}, {MacQueen}, {Marcy}, {McCullough}, {Morton}, {Narita}, {Paegert}, {Palle}, {Pepe}, {Pepper}, {Quirrenbach}, {Rinehart}, {Sasselov}, {Sato}, {Seager}, {Sozzetti}, {Stassun}, {Sullivan}, {Szentgyorgyi}, {Torres}, {Udry}, \& {Villasenor}}]{Ricker2015}
{Ricker}, G.~R., {Winn}, J.~N., {Vanderspek}, R., {et~al.} 2015, \bibinfo{title}{{Transiting Exoplanet Survey Satellite (TESS)},} Journal of Astronomical Telescopes, Instruments, and Systems, 1, 014003, \dodoi{10.1117/1.JATIS.1.1.014003}

\bibitem[{P. {Robertson} {et~al.}(2019){Robertson}, {Anderson}, {Stefansson}, {Hearty}, {Monson}, {Mahadevan}, {Blakeslee}, {Bender}, {Ninan}, {Conran}, {Levi}, {Lubar}, {Cole}, {Dykhouse}, {Kanodia}, {Nitroy}, {Smolsky}, {Tuggle}, {Blank}, {Nelson}, {Blake}, {Halverson}, {Henderson}, {Kaplan}, {Li}, {Logsdon}, {McElwain}, {Rajagopal}, {Ramsey}, {Roy}, {Schwab}, {Terrien}, \& {Wright}}]{Robertson2019}
{Robertson}, P., {Anderson}, T., {Stefansson}, G., {et~al.} 2019, \bibinfo{title}{{Ultrastable environment control for the NEID spectrometer: design and performance demonstration},} Journal of Astronomical Telescopes, Instruments, and Systems, 5, 015003, \dodoi{10.1117/1.JATIS.5.1.015003}

\bibitem[{A.~M. {Rossi} {et~al.}(2026){Rossi}, {Rainer}, {Borsa}, \& {Facchini}}]{Rossi2025}
{Rossi}, A.~M., {Rainer}, M., {Borsa}, F., \& {Facchini}, S. 2026, \bibinfo{title}{{True spin-orbit obliquity distribution: Data-driven confirmation of no clustering of misaligned planets},} \aap, 705, A142, \dodoi{10.1051/0004-6361/202555173}

\bibitem[{R. Rossiter(1924)Rossiter}]{Rossiter1924}
Rossiter, R. 1924, \bibinfo{title}{On the detection of an effect of rotation during eclipse in the velocity of the brigher component of beta Lyrae, and on the constancy of velocity of this system.,} Astrophysical Journal, 60, 15-21 (1924), 60

\bibitem[{A. {Santerne} {et~al.}(2016){Santerne}, {Moutou}, {Tsantaki}, {Bouchy}, {H{\'e}brard}, {Adibekyan}, {Almenara}, {Amard}, {Barros}, {Boisse}, {Bonomo}, {Bruno}, {Courcol}, {Deleuil}, {Demangeon}, {D{\'\i}az}, {Guillot}, {Havel}, {Montagnier}, {Rajpurohit}, {Rey}, \& {Santos}}]{Santerne2016}
{Santerne}, A., {Moutou}, C., {Tsantaki}, M., {et~al.} 2016, \bibinfo{title}{{SOPHIE velocimetry of Kepler transit candidates. XVII. The physical properties of giant exoplanets within 400 days of period},} \aap, 587, A64, \dodoi{10.1051/0004-6361/201527329}

\bibitem[{N. {Schanche} {et~al.}(2025){Schanche}, {H{\'e}brard}, {Stassun}, {Hord}, {Barkaoui}, {Bieryla}, {Ciardi}, {Collins}, {Collier Cameron}, {Hartman}, {Heidari}, {Hellier}, {Howell}, {Lendl}, {McCormac}, {McLeod}, {Parviainen}, {Radford}, {Rajpurohit}, {Relles}, {Sharma}, {Baliwal}, {Bakos}, {Barros}, {Bouchy}, {Burdanov}, {Budnikova}, {Chakaraborty}, {Clark}, {Delrez}, {Demangeon}, {D{\'\i}az}, {Donnenfield}, {Everett}, {Fukui}, {Gillon}, {Hedges}, {Higuera}, {Jehin}, {Jenkins}, {Kiefer}, {Laloum}, {Livingston}, {Lund}, {Magain}, {Maxted}, {Mireles}, {Murgas}, {Narita}, {Nikitha}, {Opitom}, {Palle}, {Patel}, {Rose}, {Sousa}, {Strakhov}, {Str{\o}m}, {Tuson}, {West}, \& {Winn}}]{Schanche2025}
{Schanche}, N., {H{\'e}brard}, G., {Stassun}, K.~G., {et~al.} 2025, \bibinfo{title}{{A Swarm of WASP Planets: Nine Giant Planets Identified by the WASP Survey},} \aj, 169, 334, \dodoi{10.3847/1538-3881/adccc6}

\bibitem[{K.~C. {Schlaufman}(2010){Schlaufman}}]{Schlaufman2010}
{Schlaufman}, K.~C. 2010, \bibinfo{title}{{Evidence of Possible Spin-orbit Misalignment Along the Line of Sight in Transiting Exoplanet Systems},} \apj, 719, 602, \dodoi{10.1088/0004-637X/719/1/602}

\bibitem[{C. {Schwab} {et~al.}(2016){Schwab}, {Rakich}, {Gong}, {Mahadevan}, {Halverson}, {Roy}, {Terrien}, {Robertson}, {Hearty}, {Levi}, {Monson}, {Wright}, {McElwain}, {Bender}, {Blake}, {St{\"u}rmer}, {Gurevich}, {Chakraborty}, \& {Ramsey}}]{Schwab2016}
{Schwab}, C., {Rakich}, A., {Gong}, Q., {et~al.} 2016, \bibinfo{title}{{Design of NEID, an extreme precision Doppler spectrograph for WIYN},} in Society of Photo-Optical Instrumentation Engineers (SPIE) Conference Series, Vol. 9908, Ground-based and Airborne Instrumentation for Astronomy VI, ed. C.~J. {Evans}, L.~{Simard}, \& H.~{Takami}, 99087H, \dodoi{10.1117/12.2234411}

\bibitem[{J.~C. {Siegel} {et~al.}(2023){Siegel}, {Winn}, \& {Albrecht}}]{Siegel2023}
{Siegel}, J.~C., {Winn}, J.~N., \& {Albrecht}, S.~H. 2023, \bibinfo{title}{{Ponderings on the Possible Preponderance of Perpendicular Planets},} \apjl, 950, L2, \dodoi{10.3847/2041-8213/acd62f}

\bibitem[{M.~M. {Sirk} {et~al.}(2018){Sirk}, {Wishnow}, {Weisfeiler}, {Jhoti}, {Curtis}, {Ishikawa}, {Finstad}, {O'Hanlon}, {Gibson}, {Edelstein}, {Halverson}, {Roy}, \& {Howard}}]{Sirk2018}
{Sirk}, M.~M., {Wishnow}, E.~H., {Weisfeiler}, M., {et~al.} 2018, \bibinfo{title}{{A optical fiber double scrambler and mechanical agitator system for the Keck planet finder spectrograph},} in Society of Photo-Optical Instrumentation Engineers (SPIE) Conference Series, Vol. 10702, Ground-based and Airborne Instrumentation for Astronomy VII, ed. C.~J. {Evans}, L.~{Simard}, \& H.~{Takami}, 107026F, \dodoi{10.1117/12.2312945}

\bibitem[{M.~F. {Skrutskie} {et~al.}(2006){Skrutskie}, {Cutri}, {Stiening}, {Weinberg}, {Schneider}, {Carpenter}, {Beichman}, {Capps}, {Chester}, {Elias}, {Huchra}, {Liebert}, {Lonsdale}, {Monet}, {Price}, {Seitzer}, {Jarrett}, {Kirkpatrick}, {Gizis}, {Howard}, {Evans}, {Fowler}, {Fullmer}, {Hurt}, {Light}, {Kopan}, {Marsh}, {McCallon}, {Tam}, {Van Dyk}, \& {Wheelock}}]{2mass}
{Skrutskie}, M.~F., {Cutri}, R.~M., {Stiening}, R., {et~al.} 2006, \bibinfo{title}{{The Two Micron All Sky Survey (2MASS)},} \aj, 131, 1163, \dodoi{10.1086/498708}

\bibitem[{D. {Souami} \& J. {Souchay}(2012){Souami} \& {Souchay}}]{Souami2012S}
{Souami}, D., \& {Souchay}, J. 2012, \bibinfo{title}{{The solar system's invariable plane},} \aap, 543, A133, \dodoi{10.1051/0004-6361/201219011}

\bibitem[{J.~S. {Speagle}(2020){Speagle}}]{dynesty2}
{Speagle}, J.~S. 2020, \bibinfo{title}{{DYNESTY: a dynamic nested sampling package for estimating Bayesian posteriors and evidences},} \mnras, 493, 3132, \dodoi{10.1093/mnras/staa278}

\bibitem[{K.~G. {Stassun} {et~al.}(2018){Stassun}, {Oelkers}, {Pepper}, {Paegert}, {De Lee}, {Torres}, {Latham}, {Charpinet}, {Dressing}, {Huber}, {Kane}, {L{\'e}pine}, {Mann}, {Muirhead}, {Rojas-Ayala}, {Silvotti}, {Fleming}, {Levine}, \& {Plavchan}}]{Stassun2018}
{Stassun}, K.~G., {Oelkers}, R.~J., {Pepper}, J., {et~al.} 2018, \bibinfo{title}{{The TESS Input Catalog and Candidate Target List},} \aj, 156, 102, \dodoi{10.3847/1538-3881/aad050}

\bibitem[{K.~G. {Stassun} {et~al.}(2019){Stassun}, {Oelkers}, {Paegert}, {Torres}, {Pepper}, {De Lee}, {Collins}, {Latham}, {Muirhead}, {Chittidi}, {Rojas-Ayala}, {Fleming}, {Rose}, {Tenenbaum}, {Ting}, {Kane}, {Barclay}, {Bean}, {Brassuer}, {Charbonneau}, {Ge}, {Lissauer}, {Mann}, {McLean}, {Mullally}, {Narita}, {Plavchan}, {Ricker}, {Sasselov}, {Seager}, {Sharma}, {Shiao}, {Sozzetti}, {Stello}, {Vanderspek}, {Wallace}, \& {Winn}}]{Stassun2019}
{Stassun}, K.~G., {Oelkers}, R.~J., {Paegert}, M., {et~al.} 2019, \bibinfo{title}{{The Revised TESS Input Catalog and Candidate Target List},} \aj, 158, 138, \dodoi{10.3847/1538-3881/ab3467}

\bibitem[{G. {Stefansson} {et~al.}(2016){Stefansson}, {Hearty}, {Robertson}, {Mahadevan}, {Anderson}, {Levi}, {Bender}, {Nelson}, {Monson}, {Blank}, {Halverson}, {Henderson}, {Ramsey}, {Roy}, {Schwab}, \& {Terrien}}]{Stefansson2016}
{Stefansson}, G., {Hearty}, F., {Robertson}, P., {et~al.} 2016, \bibinfo{title}{{A Versatile Technique to Enable Sub-milli-Kelvin Instrument Stability for Precise Radial Velocity Measurements: Tests with the Habitable-zone Planet Finder},} \apj, 833, 175, \dodoi{10.3847/1538-4357/833/2/175}

\bibitem[{G. {Stefansson} {et~al.}(2022){Stefansson}, {Mahadevan}, {Petrovich}, {Winn}, {Kanodia}, {Millholland}, {Maney}, {Ca{\~n}as}, {Wisniewski}, {Robertson}, {Ninan}, {Ford}, {Bender}, {Blake}, {Cegla}, {Cochran}, {Diddams}, {Dong}, {Endl}, {Fredrick}, {Halverson}, {Hearty}, {Hebb}, {Hirano}, {Lin}, {Logsdon}, {Lubar}, {McElwain}, {Metcalf}, {Monson}, {Rajagopal}, {Ramsey}, {Roy}, {Schwab}, {Schweiker}, {Terrien}, \& {Wright}}]{Stefansson2022}
{Stefansson}, G., {Mahadevan}, S., {Petrovich}, C., {et~al.} 2022, \bibinfo{title}{{The Warm Neptune GJ 3470b Has a Polar Orbit},} \apjl, 931, L15, \dodoi{10.3847/2041-8213/ac6e3c}

\bibitem[{J. {Tayar} {et~al.}(2022){Tayar}, {Claytor}, {Huber}, \& {van Saders}}]{Tayar2022}
{Tayar}, J., {Claytor}, Z.~R., {Huber}, D., \& {van Saders}, J. 2022, \bibinfo{title}{{A Guide to Realistic Uncertainties on the Fundamental Properties of Solar-type Exoplanet Host Stars},} \apj, 927, 31, \dodoi{10.3847/1538-4357/ac4bbc}

\bibitem[{L.~Y. {Temple} {et~al.}(2018){Temple}, {Hellier}, {Almleaky}, {Anderson}, {Bouchy}, {Brown}, {Burdanov}, {Collier Cameron}, {Delrez}, {Gillon}, {Hall}, {Jehin}, {Lendl}, {Maxted}, {Nielsen}, {Pepe}, {Pollacco}, {Queloz}, {S{\'e}gransan}, {Smalley}, {Sohy}, {Thompson}, {Triaud}, {Turner}, {Udry}, \& {West}}]{Temple2018}
{Temple}, L.~Y., {Hellier}, C., {Almleaky}, Y., {et~al.} 2018, \bibinfo{title}{{Discovery of WASP-174b: Doppler tomography of a near-grazing transit},} \mnras, 480, 5307, \dodoi{10.1093/mnras/sty2197}

\bibitem[{ {TESS Team}(2021){TESS Team}}]{TESS_2min}
{TESS Team}. 2021, TESS Light Curves - All Sectors, STScI/MAST, \dodoi{10.17909/T9-NMC8-F686}

\bibitem[{J. {Teyssandier} {et~al.}(2019){Teyssandier}, {Lai}, \& {Vick}}]{Teyssandier2019}
{Teyssandier}, J., {Lai}, D., \& {Vick}, M. 2019, \bibinfo{title}{{Formation of hot Jupiters through secular chaos and dynamical tides},} \mnras, 486, 2265, \dodoi{10.1093/mnras/stz1011}

\bibitem[{A.~H.~M.~J. {Triaud}(2018){Triaud}}]{Triaud2018}
{Triaud}, A. H.~M.~J. 2018, \bibinfo{title}{{The Rossiter-McLaughlin Effect in Exoplanet Research},} in Handbook of Exoplanets, ed. H.~J. {Deeg} \& J.~A. {Belmonte}, 2, \dodoi{10.1007/978-3-319-55333-7_2}

\bibitem[{S. {Udry} {et~al.}(2003){Udry}, {Mayor}, \& {Santos}}]{Udry2003}
{Udry}, S., {Mayor}, M., \& {Santos}, N.~C. 2003, \bibinfo{title}{{Statistical properties of exoplanets. I. The period distribution: Constraints for the migration scenario},} \aap, 407, 369, \dodoi{10.1051/0004-6361:20030843}

\bibitem[{F. {Valsecchi} {et~al.}(2015){Valsecchi}, {Rappaport}, {Rasio}, {Marchant}, \& {Rogers}}]{Valsecchi2015}
{Valsecchi}, F., {Rappaport}, S., {Rasio}, F.~A., {Marchant}, P., \& {Rogers}, L.~A. 2015, \bibinfo{title}{{Tidally-driven Roche-lobe Overflow of Hot Jupiters with MESA},} \apj, 813, 101, \dodoi{10.1088/0004-637X/813/2/101}

\bibitem[{S. {Van Der Walt} {et~al.}(2011){Van Der Walt}, {Colbert}, \& {Varoquaux}}]{numpy}
{Van Der Walt}, S., {Colbert}, S.~C., \& {Varoquaux}, G. 2011, \bibinfo{title}{{The NumPy Array: A Structure for Efficient Numerical Computation},} Computing in Science and Engineering, 13, 22, \dodoi{10.1109/MCSE.2011.37}

\bibitem[{H. {Veldhuis} {et~al.}(2025){Veldhuis}, {Espinoza-Retamal}, {Stefansson}, {Stephan}, {Martin}, {Bruijne}, {Mahadevan}, {Winn}, {Blake}, {Dai}, {Fernandes}, {Fitzmaurice}, {Ford}, {Giovinazzi}, {Gupta}, {Halverson}, {Han}, {Krolikowski}, {Ninan}, {Petrovich}, {Robertson}, {Roy}, {Schwab}, \& {Terrien}}]{Veldhuis2025}
{Veldhuis}, H., {Espinoza-Retamal}, J.~I., {Stefansson}, G., {et~al.} 2025, \bibinfo{title}{{TOI-1259Ab: A Warm Jupiter Orbiting a K-dwarf White-Dwarf Binary is on a Well-aligned Orbit},} arXiv e-prints, arXiv:2507.07737, \dodoi{10.48550/arXiv.2507.07737}

\bibitem[{P. {Virtanen} {et~al.}(2020){Virtanen}, {Gommers}, {Oliphant}, {Haberland}, {Reddy}, {Cournapeau}, {Burovski}, {Peterson}, {Weckesser}, {Bright}, {van der Walt}, {Brett}, {Wilson}, {Millman}, {Mayorov}, {Nelson}, {Jones}, {Kern}, {Larson}, {Carey}, {Polat}, {Feng}, {Moore}, {VanderPlas}, {Laxalde}, {Perktold}, {Cimrman}, {Henriksen}, {Quintero}, {Harris}, {Archibald}, {Ribeiro}, {Pedregosa}, {van Mulbregt}, \& {SciPy 1. 0 Contributors}}]{scipy}
{Virtanen}, P., {Gommers}, R., {Oliphant}, T.~E., {et~al.} 2020, \bibinfo{title}{{SciPy 1.0: fundamental algorithms for scientific computing in Python},} Nature Medicine, 17, 261, \dodoi{10.1038/s41592-019-0686-2}

\bibitem[{S. {Vissapragada} \& A. {Behmard}(2025){Vissapragada} \& {Behmard}}]{Vissapragada2025}
{Vissapragada}, S., \& {Behmard}, A. 2025, \bibinfo{title}{{The Hottest Neptunes Orbit Metal-rich Stars},} \aj, 169, 117, \dodoi{10.3847/1538-3881/ada143}

\bibitem[{S.~S. {Vogt} {et~al.}(1994){Vogt}, {Allen}, {Bigelow}, {Bresee}, {Brown}, {Cantrall}, {Conrad}, {Couture}, {Delaney}, {Epps}, {Hilyard}, {Hilyard}, {Horn}, {Jern}, {Kanto}, {Keane}, {Kibrick}, {Lewis}, {Osborne}, {Pardeilhan}, {Pfister}, {Ricketts}, {Robinson}, {Stover}, {Tucker}, {Ward}, \& {Wei}}]{Vogt1994}
{Vogt}, S.~S., {Allen}, S.~L., {Bigelow}, B.~C., {et~al.} 1994, \bibinfo{title}{{HIRES: the high-resolution echelle spectrometer on the Keck 10-m Telescope},} in Society of Photo-Optical Instrumentation Engineers (SPIE) Conference Series, Vol. 2198, Instrumentation in Astronomy VIII, ed. D.~L. {Crawford} \& E.~R. {Craine}, 362, \dodoi{10.1117/12.176725}

\bibitem[{X.-Y. {Wang} \& S. {Wang}(2026){Wang} \& {Wang}}]{WangWang2026}
{Wang}, X.-Y., \& {Wang}, S. 2026, \bibinfo{title}{{Warm Sub-Saturns Orbiting Single Stars Are Spin-Orbit Aligned},} arXiv e-prints, arXiv:2607.29558, \dodoi{10.48550/arXiv.2607.29558}

\bibitem[{X.-Y. {Wang} {et~al.}(2026{\natexlab{a}}){Wang}, {Wang}, \& {Batygin}}]{socat}
{Wang}, X.-Y., {Wang}, S., \& {Batygin}, K. 2026{\natexlab{a}}, \bibinfo{title}{{A Homogeneous Catalog of Rossiter-McLaughlin Systems: Distinct $e$-$λ$ Trends in Three Gas-Giant Mass Regimes},} arXiv e-prints, arXiv:2605.28719, \dodoi{10.48550/arXiv.2605.28719}

\bibitem[{X.-Y. {Wang} {et~al.}(2026{\natexlab{b}}){Wang}, {Wang}, \& {Ong}}]{Wang2026}
{Wang}, X.-Y., {Wang}, S., \& {Ong}, J.~M.~J. 2026{\natexlab{b}}, \bibinfo{title}{{Unified Kraft Break at {\ensuremath{\sim}}6500 K: A Newly Identified Single-star Obliquity Transition Matches the Classical Rotation Break},} \apjl, 996, L7, \dodoi{10.3847/2041-8213/ae21c5}

\bibitem[{W.~R. {Ward}(1997){Ward}}]{Ward1997}
{Ward}, W.~R. 1997, \bibinfo{title}{{Protoplanet Migration by Nebula Tides},} \icarus, 126, 261, \dodoi{10.1006/icar.1996.5647}

\bibitem[{G.~C. {Weldon} {et~al.}(2026){Weldon}, {Yee}, {Hansen}, {Naoz}, {Hartman}, {Winn}, {Butler}, {Crane}, {Evans}, {Gan}, {Howell}, {Kunimoto}, {Osip}, {Rapetti}, {Shectman}, {Stassun}, {Teske}, {Zambelli}, {Zhou}, \& {Ziegler}}]{Weldon2026}
{Weldon}, G.~C., {Yee}, S.~W., {Hansen}, B. M.~S., {et~al.} 2026, \bibinfo{title}{{Discovery of an Inflated Hot Neptune and Its Formation from Jovian Mass Loss},} arXiv e-prints, arXiv:2607.01315, \dodoi{10.48550/arXiv.2607.01315}

\bibitem[{J.~N. {Winn} {et~al.}(2010){Winn}, {Fabrycky}, {Albrecht}, \& {Johnson}}]{Winn2010}
{Winn}, J.~N., {Fabrycky}, D., {Albrecht}, S., \& {Johnson}, J.~A. 2010, \bibinfo{title}{{Hot Stars with Hot Jupiters Have High Obliquities},} \apjl, 718, L145, \dodoi{10.1088/2041-8205/718/2/L145}

\bibitem[{J.~N. {Winn} \& G. {Stef{\'a}nsson}(2025){Winn} \& {Stef{\'a}nsson}}]{Winn2025}
{Winn}, J.~N., \& {Stef{\'a}nsson}, G. 2025, \bibinfo{title}{{Orbital Decay Candidates Reconsidered: WASP-4 b Is Not Decaying and Kepler-1658 b Is Not a Planet},} \psj, 6, 300, \dodoi{10.3847/PSJ/ae21db}

\bibitem[{J.~N. {Winn} {et~al.}(2005){Winn}, {Noyes}, {Holman}, {Charbonneau}, {Ohta}, {Taruya}, {Suto}, {Narita}, {Turner}, {Johnson}, {Marcy}, {Butler}, \& {Vogt}}]{Winn2005}
{Winn}, J.~N., {Noyes}, R.~W., {Holman}, M.~J., {et~al.} 2005, \bibinfo{title}{{Measurement of Spin-Orbit Alignment in an Extrasolar Planetary System},} \apj, 631, 1215, \dodoi{10.1086/432571}

\bibitem[{J.~N. {Winn} {et~al.}(2017){Winn}, {Sanchis-Ojeda}, {Rogers}, {Petigura}, {Howard}, {Isaacson}, {Marcy}, {Schlaufman}, {Cargile}, \& {Hebb}}]{Winn2017}
{Winn}, J.~N., {Sanchis-Ojeda}, R., {Rogers}, L., {et~al.} 2017, \bibinfo{title}{{Absence of a Metallicity Effect for Ultra-short-period Planets},} \aj, 154, 60, \dodoi{10.3847/1538-3881/aa7b7c}

\bibitem[{J.-P. {Zahn}(1977){Zahn}}]{Zahn1977}
{Zahn}, J.-P. 1977, \bibinfo{title}{{Tidal friction in close binary systems.},} \aap, 57, 383

\bibitem[{J.~J. {Zanazzi} {et~al.}(2026){Zanazzi}, {MacLeod}, {Bryan}, \& {Mahadevan}}]{Zanazzi2026}
{Zanazzi}, J.~J., {MacLeod}, M., {Bryan}, M.~L., \& {Mahadevan}, S. 2026, \bibinfo{title}{{Dynamical Tides during High-Eccentricity Migration produces the Hot Jupiter Pile-up, Neptune Ridge, and Neptune Desert},} arXiv e-prints, arXiv:2606.20789, \dodoi{10.48550/arXiv.2606.20789}

\bibitem[{M. {Zechmeister} {et~al.}(2018){Zechmeister}, {Reiners}, {Amado}, {Azzaro}, {Bauer}, {B{\'e}jar}, {Caballero}, {Guenther}, {Hagen}, {Jeffers}, {Kaminski}, {K{\"u}rster}, {Launhardt}, {Montes}, {Morales}, {Quirrenbach}, {Reffert}, {Ribas}, {Seifert}, {Tal-Or}, \& {Wolthoff}}]{Zechmeister2018}
{Zechmeister}, M., {Reiners}, A., {Amado}, P.~J., {et~al.} 2018, \bibinfo{title}{{Spectrum radial velocity analyser (SERVAL). High-precision radial velocities and two alternative spectral indicators},} \aap, 609, A12, \dodoi{10.1051/0004-6361/201731483}

\bibitem[{G. {Zhou} {et~al.}(2018){Zhou}, {Rodriguez}, {Vanderburg}, {Quinn}, {Irwin}, {Huang}, {Latham}, {Bieryla}, {Esquerdo}, {Berlind}, \& {Calkins}}]{Zhou2018}
{Zhou}, G., {Rodriguez}, J.~E., {Vanderburg}, A., {et~al.} 2018, \bibinfo{title}{{The Warm Neptunes around HD 106315 Have Low Stellar Obliquities},} \aj, 156, 93, \dodoi{10.3847/1538-3881/aad085}

\end{thebibliography}
\bibliographystyle{aasjournalv7}


\end{document}